\documentclass{amsart}
\usepackage{threeparttable}
\usepackage{amsmath,amssymb,appendix,amsthm}
\usepackage{graphicx}
\usepackage{float}

 \newcommand{\eqn}[1]{Eq.~(\ref{#1})}
 
 \newcommand{\stdp}{\strut\displaystyle} 
 
\def\opex{ Opt.\ Express }

\def\prl{ Phys.\ Rev.\ Lett.\ }

\begin{document}

\title[]{Exploring laser-driven quantum phenomena from a time-frequency analysis perspective: A comprehensive study}

%

\author{Yae-lin Sheu}
\address{
1215 Hunters Glen Dr., Plainsboro, New Jersey 08536, USA\\
}

\author{Hau-tieng Wu}
\address{Department of Mathematics, University of Toronto, Toronto ON M5S 2E4, Canada. Email: hauwu@math.toronto.edu
}

\author{Liang-Yan Hsu}
\address{Department of Chemistry, Princeton University, Princeton, New Jersey 08544, USA. Email:  lianghsu@princeton.edu
}

\maketitle

\begin{abstract}
Time-frequency (TF) analysis is a powerful tool for exploring ultrafast dynamics in atoms and molecules.
While some TF methods have demonstrated their usefulness and potential in several of quantum systems, a systematic comparison among these methods is still lacking.
To this end, we compare a series of classical and contemporary TF methods by taking hydrogen atom in a strong laser field as a benchmark.
In addition, several TF methods such as Cohen class distribution other than the Wigner-Ville distribution, reassignment methods, and the empirical mode decomposition method are first introduced to exploration of ultrafast dynamics.
Among these TF methods, the synchrosqueezing transform successfully illustrates the physical mechanisms in the multiphoton ionization regime and in the tunneling ionization regime.
Furthermore, an empirical procedure to analyze an unknown complicated quantum system is provided, indicating the versatility of TF analysis as a new viable venue for exploring quantum dynamics.
\end{abstract}

\section{Introduction}

Time-dependent quantum mechanics is a fundamental topic in physics, chemistry, and engineering. Traditionally, dynamics of a quantum system such as lifetime and energy difference between states can be revealed by spectral lines in the frequency domain by using the Fourier analysis. However, the Fourier transform presents limited chronological information of a dynamical system. To explore chronological information in a quantum dynamical system, time-frequency (TF) methods are necessary.

In recent experiment advances, TF methods such as wavelet transform combined with ultrafast spectroscopy techniques were used to probe dynamics in molecules, solids and liquids \cite{Ultrafast1,Ultrafast2,Ultrafast3,Ultrafast4,Ultrafast5,Ultrafast6,Ultrafast7}, particular in light harvesting complexes\cite{Ultrafast8,Ultrafast9,Ultrafast10}.
In attosecond physics, the short-time Fourier transform (STFT), the continuous wavelet transform (CWT), Wigner-Ville distribution (WVD) (which belongs to the Cohen class distribution), were employed to uncover dynamical mechanisms including high-order harmonic generation (HHG) \cite{Atto1,Chirila,Figueira1,Antoine1,Tong_Morlet,XChu1,Chen_WVD}. Lately, Hilbert transform and the synchrosqueezing transform (SST) were applied to investigate multiple scattering \cite{Risoud} and the dynamic origin of near- and below-threshold harmonic generation, respectively \cite{Li_He,Li_Cs}. While these studies show the usefulness and potential of each individual TF analysis for probing dynamics of a quantum system, several crucial issues remain { when it comes to analysis of a quantum system with unknown complicated dynamics.}
First, different types of TF methods and the choice of window functions may provide extremely different TF representation, leading to conflicting physical interpretation.
Second, it is difficult to select a particular TF method based on previous studies, in which the physical models are different, \textit{e.g.}, 1D \cite{ Chirila,Risoud}, 2D \cite{Chirila} and 3D atoms \cite{Chirila,Tong_Morlet,XChu1,Li_He,Li_Cs}, and solved by different numerical schemes \cite{Chirila,Tong_Morlet,Li_Cs,Risoud}.
Clearly, TF representations derived from different TF methods based on different physical models cannot provide an impartial comparison.
%
Third, in the past decades, several modern TF methods, \textit{e.g.}, the Cohen class distribution, empirical mode decomposition with Hilbert spectrum (EMD-HS), reassignment methods (RM), and SST have been proposed and successfully applied to classical macroscopic dynamical systems including molecular dynamics \cite{Other1}, cardiopulmonary coupling phenomena \cite{Other2}, chronotaxic systems \cite{Other3}, lamb wave propagation \cite{Other4} and seismic data \cite{Other5}.
However, Cohen class distribution other than the WVD, EMD-HS, RM, and different forms of SST have not been discussed in quantum dynamical systems.
%
As a consequence, a comprehensive TF study in the same benchmark is essential.

To this end, we take 3D hydrogen atom as a benchmark because hydrogen is one of the most representative systems in quantum mechanics. All simulations are based on the same numerical method (time-dependent generalized pseudospectral method), and different contemporary TF analysis, including the STFT, CWT, Cohen class distributions, RM, SSTs and EMD-HS, are applied to study the simulated signals.
The systematic comparison of the TF methods enables to interpret physical processes in a quantum system.

This article is organized as follows. 
In Section \ref{Section:TheoreticalModel}, we summarize the physical model of the hydrogen system in a strong laser field and the numerical simulation.
In Section \ref{Section:TFAnalysis}, we provide an overview of the TF methods, as well as a discussion of their pros and cons.
In order to show that quantum dynamical phenomena can be depicted by the TF methods, we perform simulations with two different sets of laser parameters, including hydrogen in the regime of multiphoton ionization and tunneling ionization.
The results of TF representations by different TF methods and a discussion of their limitations and potential applications are given in Section \ref{Section:ResultDiscussion}. In Section \ref{Section:Conclusion} we conclude the paper and provide an empirical procedure to analyze an unknown complicated quantum system.

\section{Theoretical modeling}\label{Section:TheoreticalModel}

We simulate hydrogen dynamics in a linear polarized laser field in the framework of the electric dipole approximation and non-relativistic quantum mechanics.
Note that for the wavelength range of approximately $500$ to $1200$ nm, the electric dipole approximation is valid \cite{breakdown} only when the laser intensity $I_0$ is smaller than $10^{15}\sim 10^{16}\mathrm{W}/\mathrm{cm}^2$. 
The time-dependent Schr\"odinger equation for atomic hydrogen interacting with a linearly polarized field along the $z$-axis in atomic units can be expressed as 
\begin{align}
\label{TD-SE1}
i\frac{\partial \psi(\mathbf{r},t)}{\partial t}=\left( -\frac{1}{2}\nabla^2 -\frac{1}{r} - z E(t) \right) \psi(\mathbf{r},t),  
\end{align}
where $\psi(\mathbf{r},t)$ is the wave function at position $\mathbf{r}=(x,y,z)$ and at time $t$, $z=r \cos(\theta)$, and $E(t)$ is the external laser field.

To numerically solve this equation, we adopt a time-dependent generalized pseudospectral method \cite{Yao_gps, Tong_gps}
which consists of two essential steps:
(1) The spatial coordinates are optimally discretized in a nonuniform fashion by means of the generalized pseudospectral technique: the grid is denser near the origin and sparser away from the origin;
(2) A second order split operator technique in the energy representation, which allows the explicit limitation of undesirable fast-oscillating high energy components, is used to obtain an efficient and accurate time propagation of the wave function.

According to previous studies \cite{Atto2,Lewenstein}, 
dynamical phenomena such as the HHG is associated with the electric dipoles, i.e., the laser-driven electron oscillating around the stationary nucleus, 
which could be expressed as, respectively, the time-dependent induced dipole in the length form, denoted as $d_L (t)$, and in the acceleration form, denoted as $d_A (t)$ \cite{Tong_gps}:
\begin{equation}
d_L (t)=\int{\psi^*(\mathbf r,t)z\psi(\mathbf r,t){\rm d}\mathbf r}.	\label{eq.01}
\end{equation}
\begin{equation}
d_A (t)=\int{\psi^*(\mathbf r,t)\frac{{\rm d}^2z}{{\rm d}t^2}\psi(\mathbf r,t){\rm d}\mathbf r}.	\label{eq.01a}
\end{equation}

\section{Time-frequency Analysis}\label{Section:TFAnalysis}

Suppose the signal $x(t)$ is composed of finite oscillatory components, that is, $x(t)=\sum_{l=1}^L a_l(t)\cos(2\pi\phi_l(t))$, and each component has a time-varying amplitude modulation (AM), $a_l(t)>0$, and time-varying instantaneous frequency (IF) $\phi'_l(t)$, where $'$ stands for the first derivative, then the TF representation reflects the IF, which is a generalization of the notation frequency, and the localized phase information could be extracted. 
We call $a_l(t)\cos(2\pi\phi_l(t))$ an intrinsic mode type (IMT) function. More details about this kind of function could be found in Appendix A.

To disclose the time-varying nature of this kind of signal, in particular the IF and AM, the TF analysis is a powerful approach. In the following subsections we provide an overview on classical and contemporary TF methods.

\noindent\subsection{Linear TF methods}

The global analysis nature of the Fourier transform is responsible for its limitation in extracting time-varying dynamics inside an oscillatory signal. An intuitive way to resolve this limitation is analyzing the signal locally; that is, we could crop a segment of finite length and apply the Fourier transform, and expect to observe how a signal oscillates locally.
This idea leads to the STFT. We introduce a window function $g$ to crop the signal at different time and perform the Fourier analysis.
This analysis results in a time-frequency (TF) representation of the signal \cite{Book_TF_analysis}:
\begin{eqnarray}
{\mathrm {STFT}}_{x}^{g}(t,\omega) = \int_{-\infty}^{\infty} x(\zeta)g^{\ast}(\zeta-t)e^{-i\omega (\zeta-t)}{\rm d}\zeta. \label{intro.02}
\end{eqnarray}
We call ${\mathrm {STFT}}_{x}^{g}(t,\omega)$ a TF representation of the given function $x$ and $|{\mathrm {STFT}}_{x}^{g}(t,\omega)|^2$ the spectrogram of $x$.
Here the STFT by \eqn{intro.02} is different from the conventional definition by an additional modulation factor $e^{-i\omega t}$.

Nevertheless, temporal and frequency resolutions cannot be achieved simultaneously by the STFT, according to the Heisenberg uncertainty principal \cite{Book_TF_analysis}.
In other words, a wide window provides a good frequency estimation  in the STFT at the cost of poor temporal resolution, while a narrow window has the opposite trade-off. 
In this research we follow the tradition and choose the Gaussian function with a standard deviation of $\sigma$ (the full width at half maximum of the Gaussian is $2\sqrt{2\rm{ln}2}\sigma$) as the window function, i.e., $g(\zeta) = e^{\frac{-\zeta^2}{2\sigma^2}}$, which leads to the Gabor transform (GT):
\begin{eqnarray}
{\mathrm {GT}}_{x}^{g}(t,\omega) = \int_{-\infty}^{\infty} x(\zeta)e^{\frac{-(\zeta-t)^2}{2\sigma^2}}e^{-i\omega (\zeta-t)}{\rm d}\zeta. \label{intro.03}
\end{eqnarray}
Here the GT we apply here differs from the usual one by an additional modulation factor $e^{i\omega t}$.

Based on the same local analysis idea as that of the STFT, we could analyze the momentary behavior of the signal $x(t)$ by the CWT:
\begin{eqnarray}
{\mathrm {CWT}}_{x}^{g}(t,a) = \int_{-\infty}^{\infty} x(\zeta){\frac{1}{\sqrt{a}}g^{\ast}\left(\frac{\zeta-t}{a}\right)}{\rm d}\zeta, \label{intro.04}
\end{eqnarray}
where $g$ is the chosen mother wavelet and the scale parameter $a$ controls the dilation of the window, and $^*$ denote the complex conjugate \cite{Book_TF_analysis,Ten_Lectures_on_Wavelets}.
Due to the dilation nature of the transform, the TF representation analyzed by the CWT has a good time resolution and poor frequency resolution at the high frequency region and a good 
frequency resolution and poor time resolution at the low frequency region.
One of the commonly applied mother wavelet is the Morlet wavelet, and the CWT based on the Morlet wavelet is called the Morlet wavelet transform (MWT):
\begin{eqnarray}
{\mathrm {MWT}}_{x}^{g}(t,\omega) = \int_{-\infty}^{\infty} x(\zeta){\sqrt{\omega}}g^{*}(\omega(\zeta-t)){\rm d}\zeta, \label{intro.05}
\end{eqnarray}
where $\omega>0$, $g(\zeta) = \frac{1}{\sqrt\tau}e^{\frac{-\zeta^2}{2\tau^2}}e^{i\zeta}$ is the Morlet wavelet and $\tau>0$. 
The standard deviation for the dilated Morlet wavelet is $\sigma=\tau/\omega$. 
In \eqn{intro.05}, we follow the convention and use $\omega$, which is the inverse of the scale parameter $a$ in \eqn{intro.04}.
Note the fundamental difference between \eqn{intro.05} and \eqn{intro.03} -- in the GT, the Gaussian function is of a fixed width but in the MWT the width varies. 
In the MWT the window width $\sigma$ varies with $\omega$ such that the quality factor $\tau$ (the inverse of the relative bandwidth) is invariant on the TF plane \cite{Book_TF_analysis}.
In other words, the window width $\sigma$ becomes smaller as $\omega$ increases, and vice versa.
Since the frequency resolution depends on the scale, we say that the TF representation by CWT has an {\it adaptive frequency resolution}.

Both the STFT and the CWT belong to the {\it linear type {TF analysis}}, in which the signal is characterized by their inner products with a preassigned family of templates with free parameters. 
Clearly, these TF representations depend on the chosen window, which might cause artificial patterns on the analysis result.
Further, while in general the underlying structure of the signal under analysis is not known, there is no systematic way to design the window which faithfully reflects the structure. The above facts render the linear type TF analysis {\it non-adaptive} to the signal under analysis.

\noindent\subsection{Quadratic TF methods}

To resolve the non-adaptivity issue of the linear type TF analysis as well as finding the higher order structure, the
Wigner-Ville distribution (WVD) \cite{Book_TF_analysis} was proposed. 
The WVD is based on the concept of the autocorrelation function and is defined as 
\begin{eqnarray}
{\mathrm{WVD}}_{x}(t,\omega)=\int^{\infty}_{-\infty} x(t+\zeta/2){x^{*}(t-\zeta/2)} e^{-i\omega \zeta} {\rm d}\zeta. \label{intro.06}
\end{eqnarray}
Note that by a slight change of variable, we could view the WVD as a variant STFT which is free of the window choice issue. In this sense, WVD is {\it adaptive} to the signal under analysis. Also note that by a direct derivation, the spectrogram could be understood as follows \cite{Book_TF_analysis}:
\begin{align}
{|{\mathrm {STFT}}_{x}^{g}(t,\omega)|^2} 
= \,\iint_{-\infty}^{\infty}  {\mathrm{WVD}}_x(\zeta,\eta){\mathrm{WVD}}_g(\zeta-t,\eta-\omega){\rm d}\zeta{\rm d}\eta. \label{intro.02Q}
\end{align}
The WVD has several good mathematical properties, e.g., the signal energy is preserved in the WVD, 
the WVD is a real-valued function on the TF-plane and it provides a precise information about the chirp signal, like $x(t)=e^{i2\pi(\beta_0+\beta_1 t+\beta_2 t^2)}$, where $\beta_0\in\mathbb{R}$, $\beta_1>0$ and $\beta_2\in\mathbb{R}$.
 However, choosing the signal itself as the window function causes specific artifacts depending on the signal type. For example, when there are more than one oscillatory component in the signal, the interference patterns in the TF representation is inevitable, which might lead to mis-interpretation of the signal.

By construction, the WVD is quadratic in the signal which could be viewed as an energy distribution. A direct generalization of the WVD based on imposing some constraints on the {\it covariance structure} \cite{Book_TF_analysis} of the signal leads to the Cohen's class. 
In other words, the Cohen's class , which comprises all bilinear TF representations that are covariant to shifts in both time and frequency \cite{Book_TF_analysis} and has the following form:
\begin{eqnarray}
\noindent{\mathrm{C}}_{x}(t,\omega)=
\iiint x(s+\zeta/2){x^{*}(s-\zeta/2)}
e^{-i\omega\zeta}e^{-i\xi(s-t)}f(\xi,\zeta) {\rm d}\xi{\rm d}s{\rm d}\zeta, \label{intro.07}
\end{eqnarray}
where $f(\xi,\zeta)$ is an arbitrary parameter function. Different parameter functions lead to different TF representations.
Note that the WVD is a member of the Cohen's class when $f(\xi,\zeta) = 1$. Different parameter functions lead to different TF techniques.

One particular technique is eliminating the interferences in the WVD when there are more than one oscillatory component. 
To be more specific, we can employ a separable parameter function $f(\xi,\zeta)=G(\xi)h(\zeta)\equiv{\mathrm{W}}_{h}(\xi,\zeta)$ in \eqn{intro.07}, where suitably chosen $G(\xi)$ and $h(\zeta)$ permit a continuous and independent control of the interferences in time and frequency, respectively. 
The corresponding representation is called the {\it smoothed pseudo-Wigner-Ville distribution} (SPWVD) \cite{Book_TF_analysis} and could be rewritten as 
\begin{align}
{\mathrm{SPWVD}}_{x}^{g,h}(t,\omega)
=&\int^{\infty}_{-\infty}g(t-s)\int^{\infty}_{-\infty}h(\zeta)x(s+\zeta/2){x^{*}(s-\zeta/2)} e^{-i\omega \zeta} {\rm d}\zeta{\rm d}s \label{spwd.01} \\
=& \iint^{\infty}_{-\infty}{\mathrm{WVD}}_{x}(s,\xi){\mathrm{W}}_{h}(s-t,\xi-\omega){\rm d}s{\rm d}\xi, \label{spwd.02}
\end{align}
where $g(\zeta)$ is the inverse Fourier transform of $G(\xi)$.
Particularly, $G(\xi)$ and $h(\zeta)$ are even functions with $g(0)=1$ and $H(0)=1$, where $H(\xi)$ is the Fourier transform of $h(\zeta)$.

We mention that we could consider a variation of the Cohen's class to get the family of the {\it {affine class}} \cite{Book_TF_analysis}, e.g., the affine WVD or the affine SPWVD. 
The transform considered in the affine class is the time-scale covariant and the TF representation also has the adaptive resolution feature. To simplify the discussion, we do not study the affine class transforms, but we mention that their performance is similar to those in the Cohen's Class. 
The above-mentioned transforms are overall called the {\it quadratic TF analysis}.
A higher-order generalization is also possible, and we refer the reader of interest to, for example, \cite{oNeill,Boashash}.


\noindent\subsection{Reassignment Methods}

As features of a TF representation computed by a linear type of TF method are smeared by the introduced window functions and those by the quadratic type ones are obscured by interferences, the RM was proposed to sharpen the resolution of the TF representation.
Generally speaking, the coefficients of the TF representation at $(t,\omega)$ is reallocated to a different point $(\hat{t},\hat{\omega})$ according to a predefined reallocation rule \cite{Ch5, rmAuger,Auger2}.
A common choice of the reallocation rule is to assign values of a TF representation to the local centroids.
Note that this is different from the averaging idea behind the STFT (\eqn{intro.02Q}).
In this study we apply the RM to the STFT and SPWVD, and called the methods RM-STFT and RM-SPWVD, respectively. 

The reallocation rule for the RM-STFT is derived by estimating the local centers of gravity, denoted as:
\begin{align}
\hat{t}_{x}(t,\omega)=\left\{\begin{array}{ll}
 \stdp  \frac{\Re\left\{{{\mathrm {STFT}}_{x}^{tg}(t,\eta)}({{\mathrm {STFT}}_{x}^{g}(t,\eta)})^{\ast}\right\}}{|{{\mathrm {STFT}}_{x}^{g}(t,\eta)}|^2} &\mbox{, when}\,\, {\mathrm {STFT}}_{x}^{g}(t,\eta)\ne0\\
\infty & \mbox{, when}\,\, {\mathrm {STFT}}_{x}^{g}(t,\eta)=0.\end{array}\right.	\label{reassign.02b}
\end{align}
\begin{align}
\hat{\omega}_{x}(t,\omega)=\left\{\begin{array}{ll}
 \stdp  \frac{-\Im\left\{{{\mathrm {STFT}}_{x}^{{\rm d}g}(t,\eta)}({{\mathrm {STFT}}_{x}^{g}(t,\eta)})^{\ast}\right\}}{|{{\mathrm {STFT}}_{x}^{g}(t,\eta)}|^2} &\mbox{, when}\,\, {\mathrm {STFT}}_{x}^{g}(t,\eta)\ne0\\
\infty & \mbox{, when}\,\, {\mathrm {STFT}}_{x}^{g}(t,\eta)=0.\end{array}\right.	\label{reassign.03b}
\end{align}

\noindent
The notations $tg$ and ${\rm d}g$ stand for the window function $tg(t)$ and the first derivative of $g(t)$ in the STFT, respectively.

The RM-STFT is thus defined as
\begin{align}
{\mbox{RM-STFT}}^{g}_{x}(\hat{t},\hat{\omega}) 
= \,\iint^{\infty}_{-\infty}|{\mathrm{STFT}_{x}^{g}}(t,\omega)|^2\delta(\hat{t}_x(t,\omega)-\hat{t},\hat{\omega}_x(t,\omega)-\hat{\omega}){\rm d}t{\rm d}\omega. \label{reassign.04}
\end{align}
Essentially, this formula reassign the energetic contents of the spectrogram to the new location $(\hat{t}_x,\hat{\omega}_x)$.
As a consequence, the RM leads to a substantially improved resolution in the TF representation.
Note that the reassignment rules \eqn{reassign.02b} and \eqn{reassign.03b} can lead to the group delay and the IF of the bandpass filtered signal $y(t)={\mathrm{STFT}_{x}^{g}}(t,\omega)$ \cite{Auger2}, using only the unwrapped phase of the ${\mathrm{STFT}_{x}^{g}}(t,\omega)$.
However these physical meaningful expressions are numerically inefficient.
Other forms of reassignment operators can be found in \cite{Book_TF_analysis}.

Next we consider SPWVD. Although the SPWVD can smooth out the interferences in the WVD, the smoothing functions introduce artificial broadening. Applying the reassignment technique on the TF representation of the SPWVD can reduce such artifacts. Similar to \eqn{reassign.04}, the reassigned representation of the SPWVD is
\begin{align}
{\mbox{RM-SPWVD}}_{x}(\hat{t},\hat{\omega}) 
=\, \iint^{\infty}_{-\infty}{\mathrm{SPWVD}_{x}^{g}}(t,\omega)\delta(\hat{t}_x(t,\omega)-\hat{t},\hat{\omega}_x(t,\omega)-\hat{\omega}){\rm d}t{\rm d}\omega. \label{reassign.05}
\end{align} 
where the reassignment rule $(\hat{t}_{x},\hat{\omega}_{x})$ is determined by the concept of expectation:
\begin{align} 
\hat{t}_{x}(t,\omega)&=\iint^{\infty}_{-\infty}s{\mathrm{WVD}}_{x}(s,\xi){\mathrm{W}}_{h}(s-t,\xi-\omega){\rm d}s{\rm d}\xi \label{reassign.06} \\
\hat{\omega}_{x}(t,\omega)&=\iint^{\infty}_{-\infty}\xi{\mathrm{WVD}}_{x}(s,\xi){\mathrm{W}}_{h}(s-t,\xi-\omega){\rm d}s{\rm d}\xi. \label{reassign.07} 
\end {align}
Despite RMs are intuitive techniques to sharpen the linear type and quadratic type TF representations, inverse routines as well as mode reconstruction are not available.

\noindent\subsection{Synchrosqueezing transform}

In this section we describe the SST, which is a special RM aiming to address the intrinsic blurring issue in the linear TF methods.
{To be more precise,} the SST manifests IF characteristic according to the reallocation rule that consists of solely the frequency information, rather than the centroid of the TF representation \cite{SST1,SST2}.
In addition to sharpening the TF representation, the causality property of the signal is preserved in the SST, which allows the decomposition of oscillatory components when the signal is composed of several oscillatory components.
We refer the reader who has interest in the theoretical analysis {and other details, like reconstructing each oscillatory component and robust to noise,} of SST to \cite{SST1, SST2, SST3}.
In this section we describe the {synchrosqueezed} STFT (SST-STFT) and the {synchrosqueezed} CWT (SST-CWT).

\subsubsection{SST-STFT}

Consider a ${\mathrm {STFT}}_{x}^{g}(t,\omega)$ of $x(t)$ with a window function $g(t)$ such that {$\mbox{supp}(G(\omega))\subset\left[-\frac{d}{2},\frac{d}{2}\right]$}, where $G(\omega)$ is the Fourier transform of $g$ and $d$ is the smallest gap of IFs of any two consecutive oscillatory components.
The SST-STFT is given as
\begin{align}
{\mbox{SST-STFT}}_{x}^{g}(t,\omega) 
=\int^{\infty}_{-\infty} {\mathrm {STFT}}_{x}^{g}(t,\eta)\frac{1}{\alpha\sqrt{\pi}}e^{-\frac{{|\omega-\hat{\omega}_x (t,\eta)|}^2}{\alpha}} \,{\rm d}\eta, \label{sst_STFT.03} 
\end{align}
where $\hat{\omega}_x (t,\eta)$ is the reallocation rule and $\alpha>0$ is a controllable smoothing parameter for the resolution, which in practice is chosen to be small.
The reallocation rule given by the following equation utilizes the phase information hidden inside the smeared TF representation:
\begin{align}
\hat{\omega}_x (t,\eta)=\left\{\begin{array}{ll}
 \stdp\frac{-i\partial_t{\mathrm {STFT}}_{x}^{g}(t,\eta)}{{\mbox {STFT}}_{x}^{g}(t,\eta)} &\mbox{when}\,\, {\mathrm{STFT}}_{x}^{g}(t,\eta)\ne0\\
 \infty & \mbox{when}\,\,{\mathrm {STFT}}_{x}^{g}(t,\eta)=0.\end{array}\right.	\label{sst_STFT.04}
\end{align}
Note that only the frequency reassignment operator is considered in the SST-STFT, so that the causality property of the signal can be preserved.
A slight modification of $\hat{\omega} (t,\eta)$ based on the second-order information of the IF \cite{mSST}
might further improve the TF representation.
Indeed, when the interested oscillatory component could be approximated by a chirp function \cite{mSST,rmAuger}, although the TF representation is sharpened by the SST-STFT, a mild spreading is inevitable. We call this mild spreading the {\it diffusive pattern}. To cope with the diffusive pattern, we could consider the following reassignment rule: 
\begin{eqnarray}
\check{\omega}_x (t,\eta)=  
\left\{\begin{array}{ll}
 \stdp\hat{\omega}_x (t,\eta) + c(t,\eta)(t-\hat{t}_x(t,\eta)) &{\mbox{when}}\,\, {\partial_{\omega} \hat{t}_x}(t,\eta)\ne0\\
     {\hat{\omega}_x} (t,\eta) & \mbox{otherwise},\end{array}\right.	 \label{sst_STFT.06}
\end{eqnarray}
where $\hat{t}_x(t,\eta)$ is defined as
\begin{eqnarray}
{{\hat t_x}(t,\eta) = t + i\frac{\partial_\omega{\mathrm {STFT}}_{x}^{g}(t,\eta)}{{\mathrm {STFT}}_{x}^{g}(t,\eta)}\quad \mbox{and}\quad c(t,\eta)=\frac{\partial_t{\hat{\omega}_x} (t,\eta)}{\partial_\omega \hat{t}_x (t,\eta)}.} \label{sst_STFT.08}
\end{eqnarray}
The reconstruction formulas \cite{SST2} for each IMT function $x_k(t)$ from ${\mathrm {SST}}_{x}^{g,\mathrm{STFT}}(t,\omega)$ is
\begin{align}
x_k(t)=\Re\left\{ {C_{g}^{-1}}\int^{{\omega_k(t)}+\frac{\epsilon}{2}}_{{\omega_k(t)}-\frac{\epsilon}{2}} {\mbox {SST-STFT}}_{x}^{g}(t,\omega){\rm d}\omega   \right\}   \label{sst_STFT.05}
\end{align}
where $\epsilon\ll1$, $\Re$ denotes taking the real part, and $C_{g}\equiv g(0)$.

\subsubsection{SST-CWT}

The definition of the SST-CWT is similar to that of the SST-STFT. For a CWT with a mother wavelet $g$ such that $\mbox{supp}(G(\omega))\subset\left[1-\Delta,1+\Delta\right]$, with $\Delta<\frac{d}{1+d}$, the SST-CWT is given by 
\begin{align}
{\mbox {SST-CWT}}_{x}^{g}(t,\omega)
=\,\int^{\infty}_{-\infty} \eta^{-3/2}{\mathrm {CWT}}_{x}^{g}(t,\eta)\frac{1}{\alpha\sqrt{\pi}}e^{-\frac{{|\omega-\hat{\omega}_x (t,\eta)|}^2}{\alpha}} \,{\rm d}\eta,\label{sst_CWT.01}
\end{align}
where $\alpha>0$ is a smoothing parameter and $\hat{\omega}_x (t,\eta)$ is the reallocation rule defined by
\begin{align}
\hat{\omega}_x (t,\eta)=\left\{\begin{array}{ll}
 \stdp\frac{-i\partial_t{\mathrm {CWT}}_{x}^{g}(t,\eta)}{{\mathrm {CWT}}_{x}^{g}(t,\eta)} &\mbox{when}\,\, {\mathrm {CWT}}_{x}^{g}(t,\eta)\ne0\\
 \infty & \mbox{when}\,\, {\mathrm {CWT}}_{x}^{g}(t,\eta)=0.\end{array}\right.	\label{sst_CWT.02}
\end{align}

When describing an oscillatory component with a fast varying IF, the TF representation in SST-CWT may show a slight spreading. While the second-order information of the IF has been discussed in the SST-STFT\cite{mSST}, it is not yet been studied. 

The reconstruction formula \cite{SST2, SST3} for each IMT function $x_k(t)$ under SST-CWT is 
\begin{align}
x_k(t)=\Re\left\{ R_{g}^{-1}\int^{{\omega_k(t)}+\frac{\epsilon}{2}}_{{\omega_k(t)}-\frac{\epsilon}{2}} a^{-3/2}{\mbox {SST-CWT}}_{x}^{g}(t,a){\rm d}a   \right\}   \label{sst_CWT.03}
\end{align}
where $\epsilon\ll1$, and $R_{g}:=\int^{\infty}_{-\infty}{\frac{G(\eta)}{\eta}{\rm d}\eta}$.

The results of the SST are ``adaptive'' to the signal in the sense that the error in the estimation depends only on the first three moments of the mother wavelet instead of the profile of the mother wavelet \cite{SST1,SST2,SST3}. In other words, the influence of the chosen window on the associated TF representation is minimized compared with the linear TF methods. 
In addition to the above properties, the SST is robust to several different kinds of noise, which might be slightly non-stationary \cite{SST3}. {An important property shared by the SST-STFT, SST-CWT and RM-STFT is that by taking a short window, the fast varying IF could be well captured.}

At the first glance, the results of the reassignment technique and SST seem to break the well-known Heisenberg uncertain principle, which says that the temporal and frequency resolution cannot be achieved simultaneously. However, in the reassignment technique and SST, we have shown that we could obtain a TF representation with an almost perfect time and frequency resolution. The main reason is that we focus on the oscillatory signals, in particular the adaptive harmonic model but not the whole $L^2$ space. We mention that the behavior of the RM and SST on the general $L^2$ function is an open question.

\noindent\subsection{EMD}

One popular way to define the IF of a given signal $x(t)$ is via finding the analytic representation of $x(t)$ with the aid of the Hilbert transform:
\begin{eqnarray}
\tilde{x}(t)= x(t) + i\mathcal{H}(x(t)) = a(t)e^{i\Phi(t)},  \label{emd.01}
\end{eqnarray}
where $\mathcal{H}(\tilde{x(t)})$ denotes the Hilbert transform of $x(t)$. In this equation, $a(t)$ and $\Phi(t)$ are the modulus and phase of $\tilde x(t)$, respectively \cite{Cohen,DGabor}.
The IF of $x(t)$ is thus defined as the rate of the varying phase:
\begin{eqnarray}
 \omega(t)=\frac{1}{2\pi}\frac{\rm{d}\Phi(t)}{{\rm d}t}.   \label{emd.02}
\end{eqnarray}
However, the IF provided by \eqn{emd.02} is not always meaningful.
To deal with this problem, the empirical mode decomposition algorithm (EMD) \cite{hht}
was proposed to decompose $x(t)$ into a series of functions called the intrinsic mode function (IMF), $x=\sum_{k=1}^K x_k(t)$,
on which the Hilbert transform can be applied subsequently, via an algorithm called the sifting process.
IMFs must satisfy the following two conditions: (i) In the whole data set of a signal, the number of extrema and the number of zero crossings must either be equal or differ at most by one; (narrow band)
(ii) At any point, the mean value of the envelope defined by the local maxima and the envelope defined by the local minima is zero (adoption of local properties). 
Note that in general an IMF is different from an IMT function.
Details of the EMD algorithm could be found in, for example, \cite{hht}. 

For each IMF $x_k(t)$, the corresponding IF and AM amplitudes can be estimated by \eqn{emd.01} via the Hilbert transform and \eqn{emd.02}.
By assigning the IF and amplitudes of all IMFs on a TF plane, we obtain the Hilbert spectrum (HS) $\rm{HS}_{x}(t,\omega)$ (HS), which is a TF representation for $x(t)$ determined by
\begin{eqnarray}
{\mathrm{HS}}_{x}(t,\omega)= \sum_{k}a_k(t)\delta(\omega-\omega_k(t)),  \label{intro.08}
\end{eqnarray}
\noindent where $\delta(\omega)$ denotes the Dirac delta measure. 
Although the EMD along with HS has been applied to several fields, due to a number of heuristic and {\it ad hoc} elements in the EMD algorithm, it is difficult to analyze its accuracy and limitation.
In addition, despite the solid mathematical support, the Hilbert transform might be limited when applied to analyze the momentary dynamics of an oscillatory signal. First, due to the slow decay nature of the kernel in the Hilbert transform, keeping the causality of the signal structure might be difficult. Second, when the signal has time-varying amplitude and frequency, in general there is no guarantee to get the correct analytic signal from a given real oscillatory signal by the Hilbert transform in \eqn{emd.01}.

We mention that the part of using the Hilbert transform to obtain the AM and IF of each IMF could be replaced by using the SST-STFT or SST-CWT. In this case, the method is regarded as the EMD-SST-STFT and EMD-SST-CWT.


A comparison of different TF methods discussed in this study is summarized in Table 1.


\begin{table}\label{list1}
\centering
\textbf{Table~1} 
Summary of TF methods in This Study\\[1ex]
\begin{tabular}{lllll}
Type      & Multiresolution  & Choice of   & Inverse    &  Artifacts in TF Representation\\
          &                  & Parameters  & Transform  &  \\
\hline
STFT      & No        & Yes     & Yes       & broadening by the window function          \\
CWT       & Yes       & Yes     & Yes       & broadening by the window function          \\
WVD       & No        & No   & Yes  & interference patterns   \\
SPWVD     & No        & Yes & Yes  & broadening by the filter function \\
RM-STFT   & No        & Yes     & No  & {causality is not preserved}\\
RM-SPWVD  & No        & Yes     & No  & {causality is not preserved}\\
SST-STFT  & No        & Yes     & Yes       & {Diffusive pattern} in the fast-varying IF      	\\
$2$nd order SST-STFT & No  & Yes& Yes       &                                     \\ 		
SST-CWT   & Yes       & Yes     & Yes       & {Diffusive pattern} in the fast-varying IF      	\\		
EMD-HS    & Yes? $\dagger$       & Yes $\ddagger$      & Yes/No $\diamond$ & mode mixing, {etc.${}^*$} \\
EMD-SST-STFT    & Yes? $\dagger$       & Yes $\ddagger$      & Yes/No $\diamond$ & mode mixing \\

\end{tabular}
\begin{tablenotes}
\item [] {{$\dagger$: The multiresolution-like behavior of the sifting process was studied in \cite{emd_Patrick}. $\ddagger$: The stopping criteria of the sifting process depends on tuning several parameters. {$\diamond$: The inversion from the HS is not possible for the definition (\ref{intro.08}). It is possible if we define the HS as $\sum_{k}a_k(t)e^{i\Phi_k(t)}\delta(\omega-\omega_k(t))$.} ${}^*$: It is sensitive to noise; a shortly existing oscillatory component destroys the whole analysis.} }
\end{tablenotes}
\end{table}


\section{Results and Discussions}\label{Section:ResultDiscussion}

In this section we employ the aforementioned TF methods on the time-dependent dipole computed by the time-dependent generalized pseudospectral method at the {\itshape ab initio} level.
In a strong laser field, a variety of dynamical processes can occur in atoms and molecules, such as ponderomotive effect, the AC Stark effect, HHG \cite{SHLin3}, multiphoton ionization\cite{SHLin1,SHLin2,Fuji1}, above threshold ionization, tunneling ionization, rescattering of electron wavepacket \cite{Gavrila, Gibbon}, {\itshape etc}.
Up to date, the STFT \cite{Atto1,Chirila,Figueira1}, CWT \cite{Antoine1,XChu1,Tong_Morlet}, WVD \cite{Chen_WVD}, and Hilbert transform \cite{Risoud} have been engaged in investigation of estimation of emission time and multiple scattering.
However, to the best of our understanding there is no comprehensive study comparing these TF methods in a quantum dynamical system.

By taking the well-studied atomic hydrogen as a benchmark, here we focus on the laser-driven hydrogen in the multiphoton ionization regime and in the tunneling ionization regime. 
When the laser photon energy $\hbar\omega$ is much smaller than the ionization potential $\rm{I_p}$, the multiphoton ionization and the tunneling ionization can be described by a dimensionless Keldysh parameter $\gamma_K=\sqrt{{\rm{I_p}}/{2U_P}}$, where $U_P=E_0^2/{4\omega_0^2}$ is the ponderomotive potential \cite{Atto2,SHLin2}.
Similar physical dynamics can be achieved for any atom-field interaction given a fixed $\gamma_K$ \cite{Power}.
Generally speaking, multiphoton ionization of atoms become dominant when $\gamma_K\gg1$, while tunneling ionization is predominant when $\gamma_K\ll1$ \cite{Atto2}.
In the following we perform simulations with laser parameters in these two regimes.

\subsection{Multiphoton Ionization Regime}

\begin{figure*}
	  \includegraphics[width=0.495\hsize]{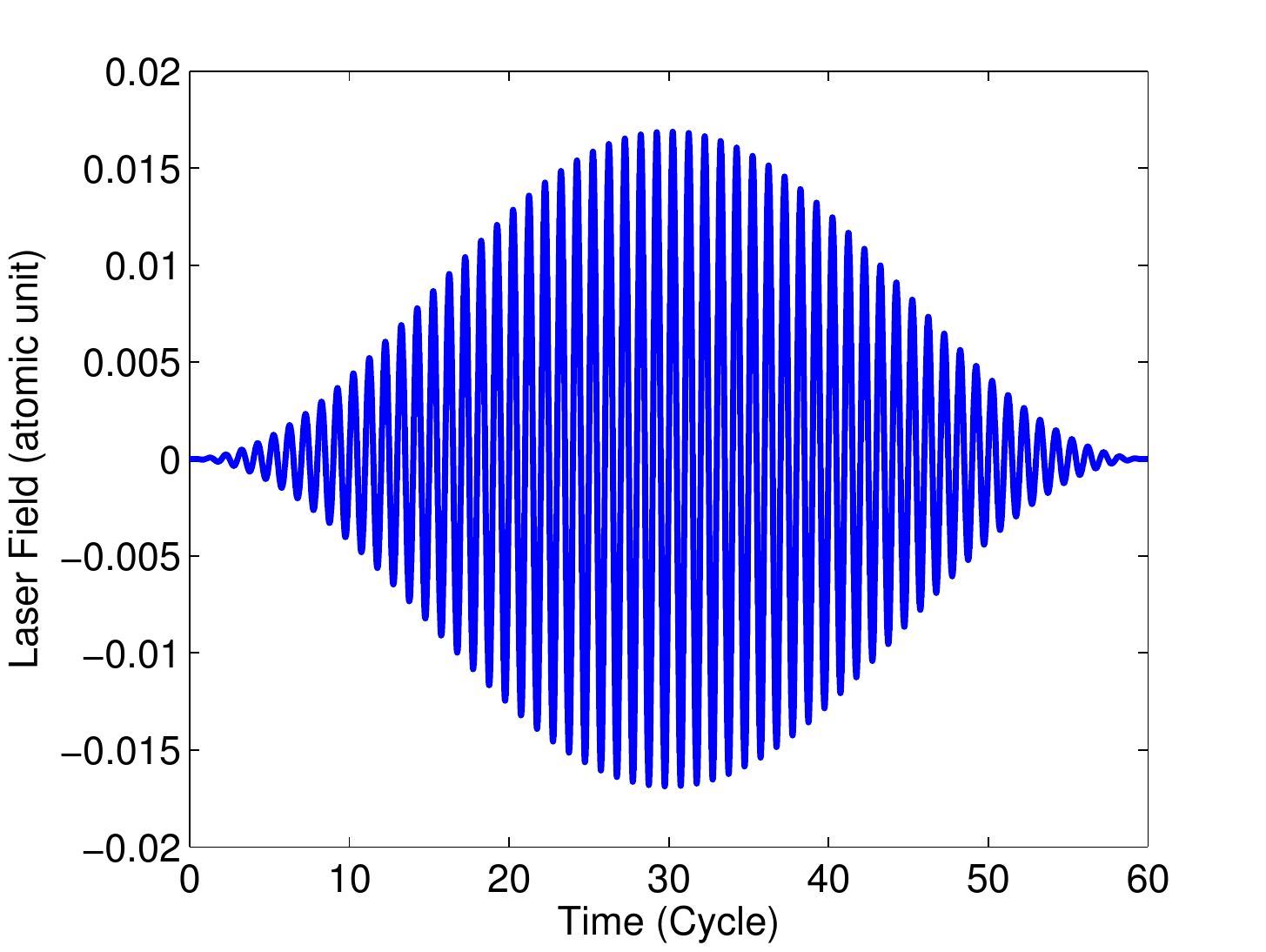}
		\includegraphics[width=0.495\hsize]{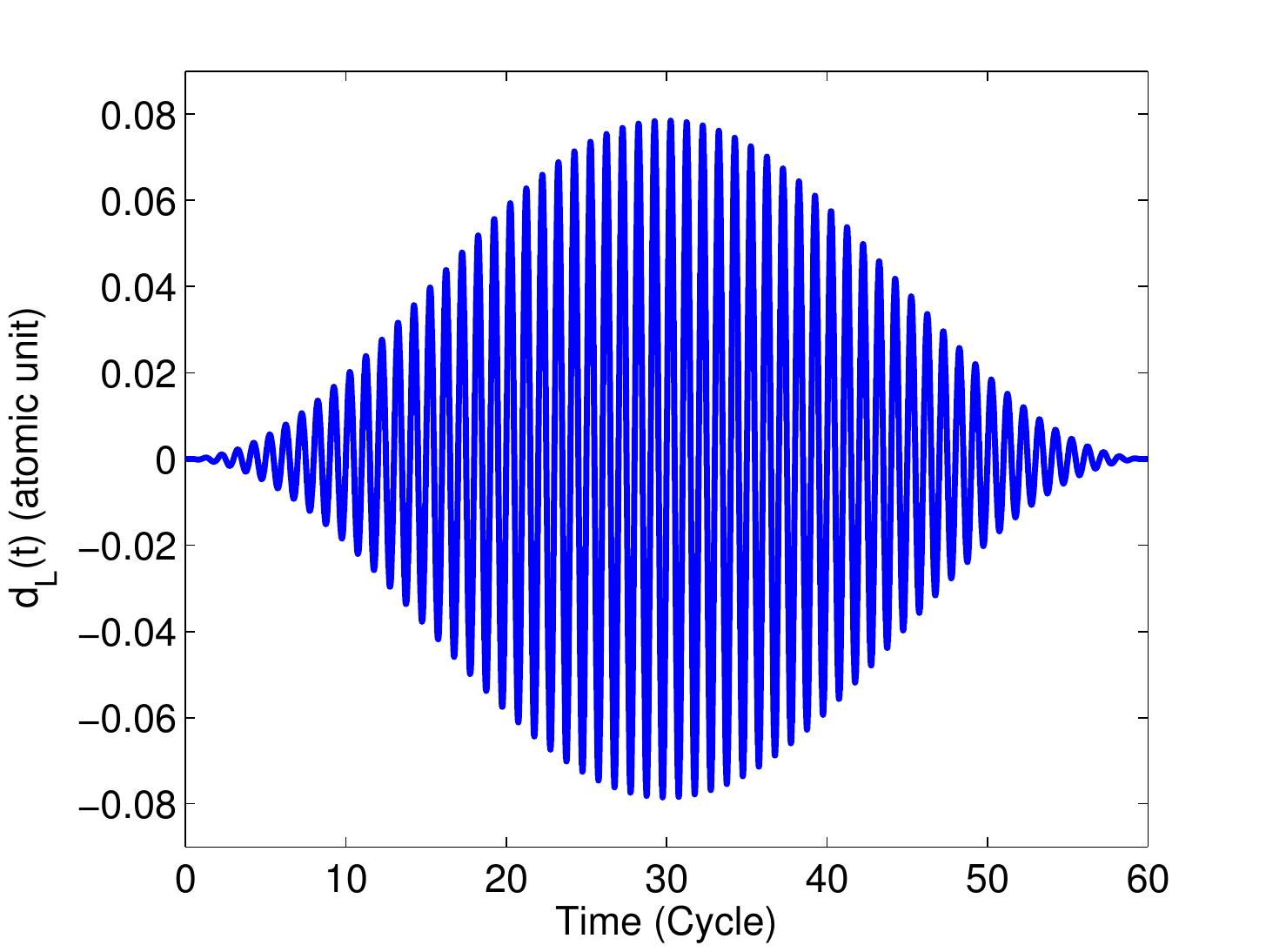}
		\hspace*{3.0cm}(a)\hspace{6.5cm}(b)\\
    \includegraphics[width=0.495\hsize]{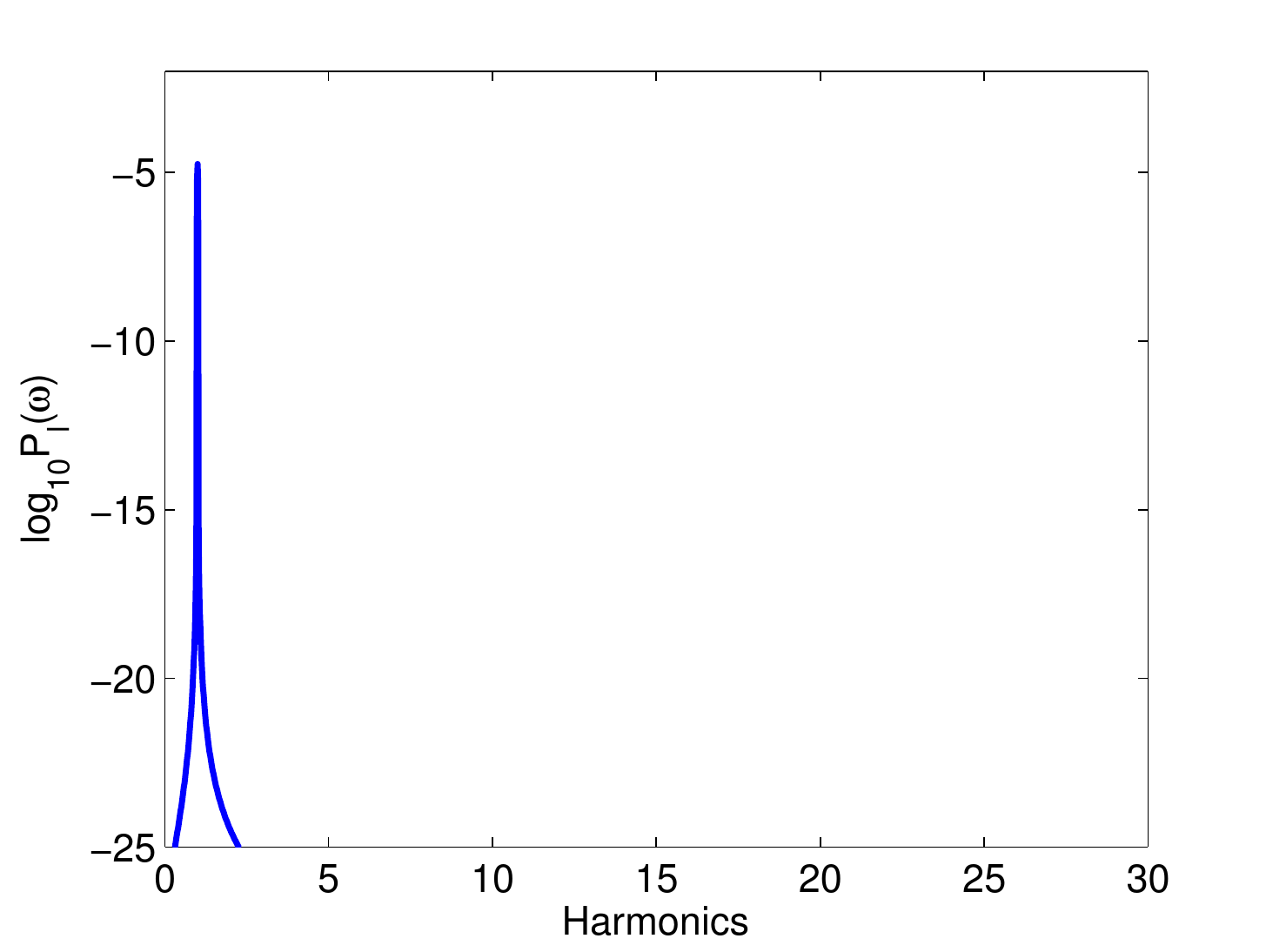}
		\includegraphics[width=0.495\hsize]{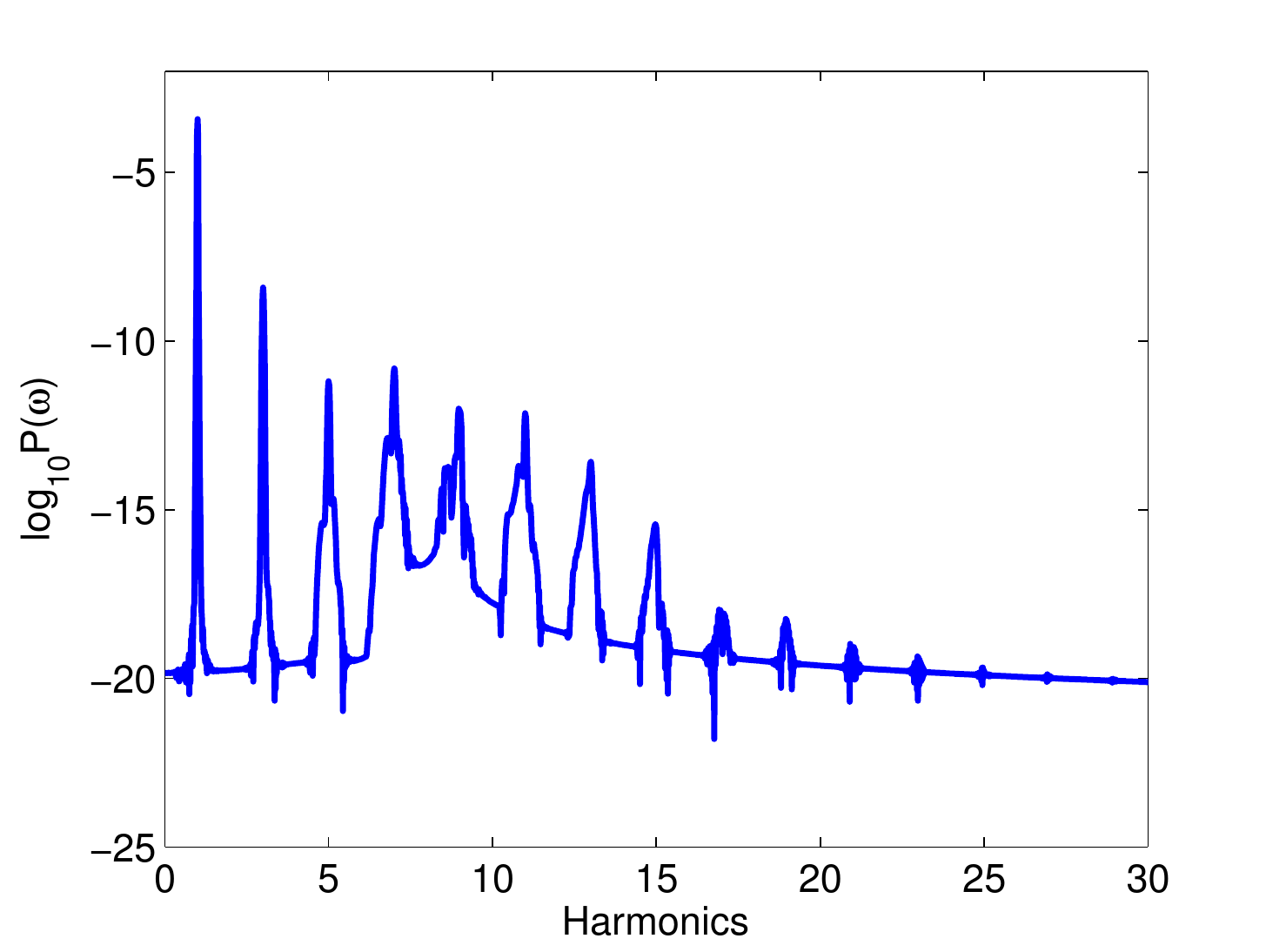}
		\hspace*{3.0cm}(c)\hspace{6.5cm}(d)\\
\vspace{-.4cm} \caption{ The simulation of laser-driven hydrogen in the multiphoton ionization regime. The laser wavelength is $880$ nm, and the laser intensity of $I_0=10^{13}\;\mbox{W/cm}^2$, corresponding to the Keldysh parameter of $\gamma_K=3.07$. (a) The laser profile. (b) The induced dipole moment $d_L(t)$. (c) The power spectrum of the laser profile. (d) The power spectrum of $d_L(t)$, respectively. Note that the laser profile and $d_L(t)$ are very similar, yet very different in their spectral components.
 }
  \label{Fig1}
\end{figure*}

In the first simulation, the laser field parameters are arranged such that the Keldysh parameter is $\gamma_K=3.07$, indicating that multiphoton ionization is the major mechanism.
The laser wavelength is $880$ nm, which corresponds to $\omega_0\approx0.05178273$ in the atomic unit (a.u.), 
and the laser intensity of $I_0=10^{13}\;\mbox{W/cm}^2$, which corresponds to a field amplitude $E_0\approx0.0169$ (a.u.).  
Note that for this $\gamma_K$ value, both tunneling and multiphoton ionization can occur, but the latter is dominant.
The laser field (Fig.~\ref{Fig1}(a)) $E(t)=E_0F(t)\sin(\omega_0t)$ has a profile of $F(t)=\sin^2 (\pi t/(nT))$, where $n=60$ is the pulse length measured in the optical cycle ($T=2\pi/\omega_0$). 

The computed induced dipole in the length form $d_L(t)$ is presented in Fig.~\ref{Fig1}(b). The power spectrum \cite{Tong_gps}, computed by the squared Fourier spectrum, of the laser field and $d_L(t)$ are shown in Fig.~\ref{Fig1}(c)and ~\ref{Fig1}(d), respectively. 
Both the length and the acceleration forms of dipole moment present the same detail structures in their power spectra \cite{Tong_gps}. Here we present the analysis using $d_L(t)$ for the multiphoton ionization case.

Although the profiles of $d_L(t)$ and the applied laser field look similar in the time domain, they are different in the frequency domain. 
For example, while the power spectrum of the laser field has only one peak located at $\omega_0$, that of $d_L(t)$ reveals odd harmonics due to the parity symmetry \cite{Atto2}. However, the meaning of the substructures within the odd harmonics and their corresponding dynamics are unclear.

\begin{figure*}
	  \includegraphics[width=0.495\hsize]{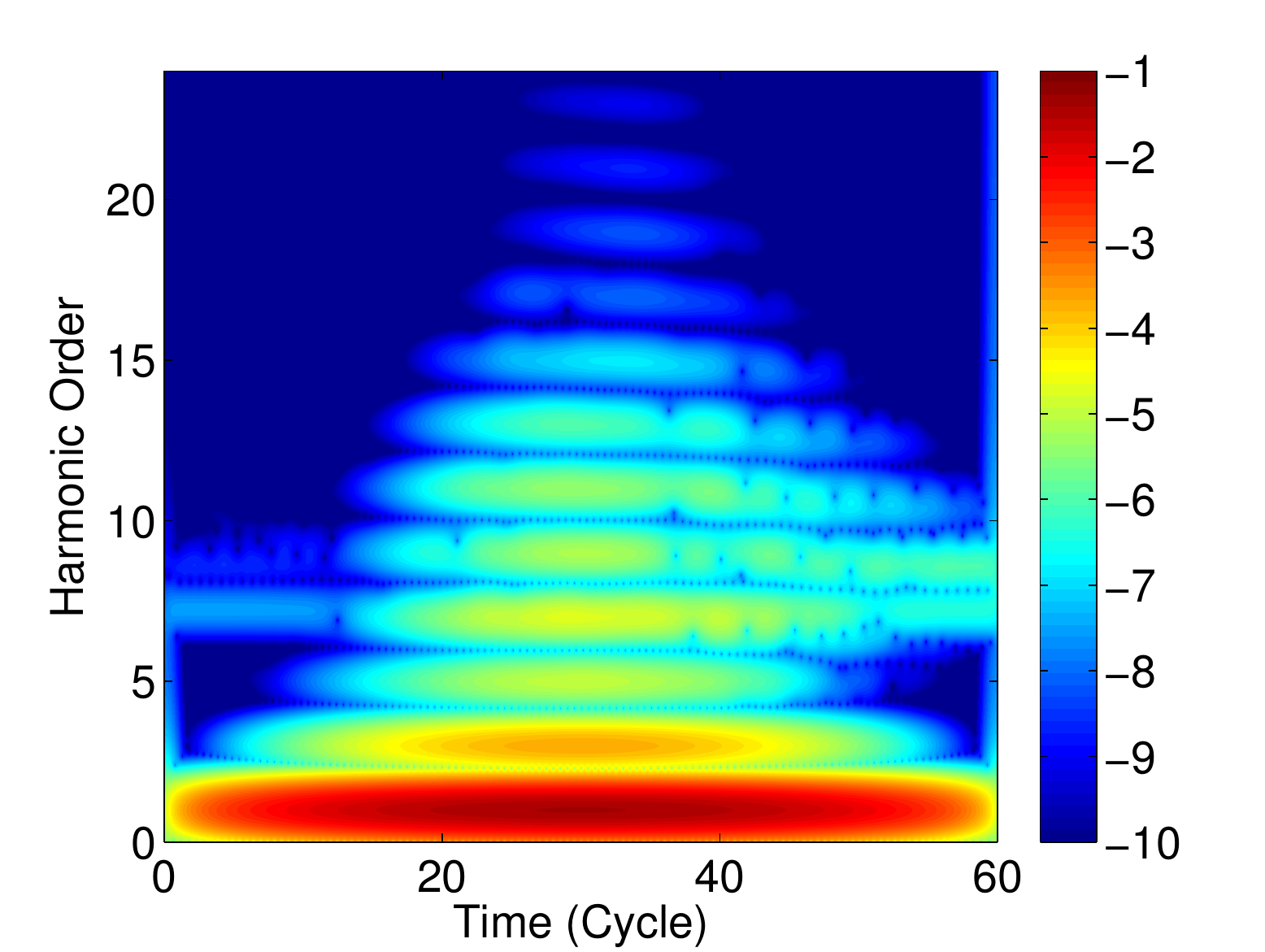}
		\includegraphics[width=0.495\hsize]{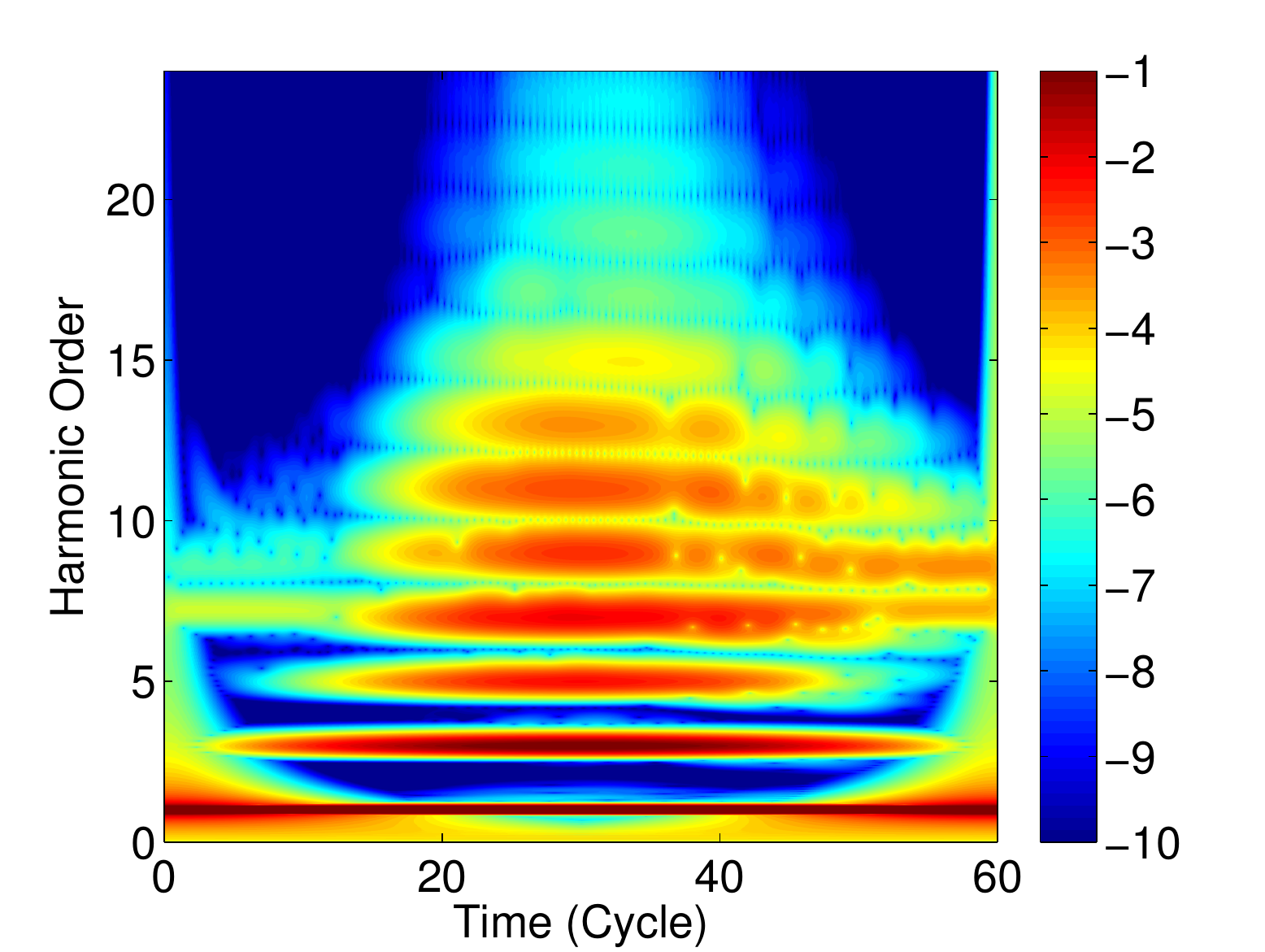}
		\hspace*{3.0cm}(a)\hspace{6.5cm}(b)\\
    \includegraphics[width=0.495\hsize]{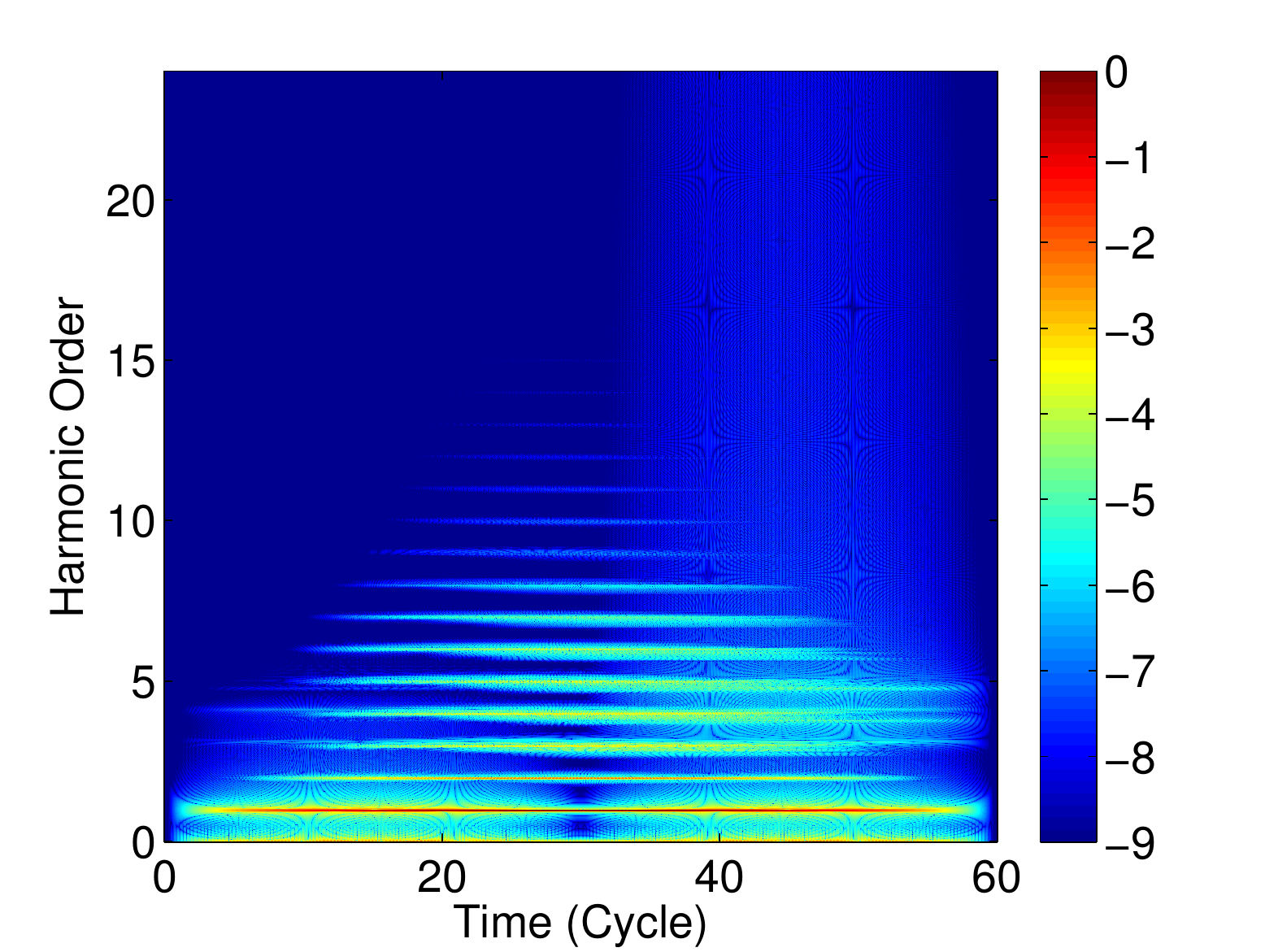}
		\includegraphics[width=0.495\hsize]{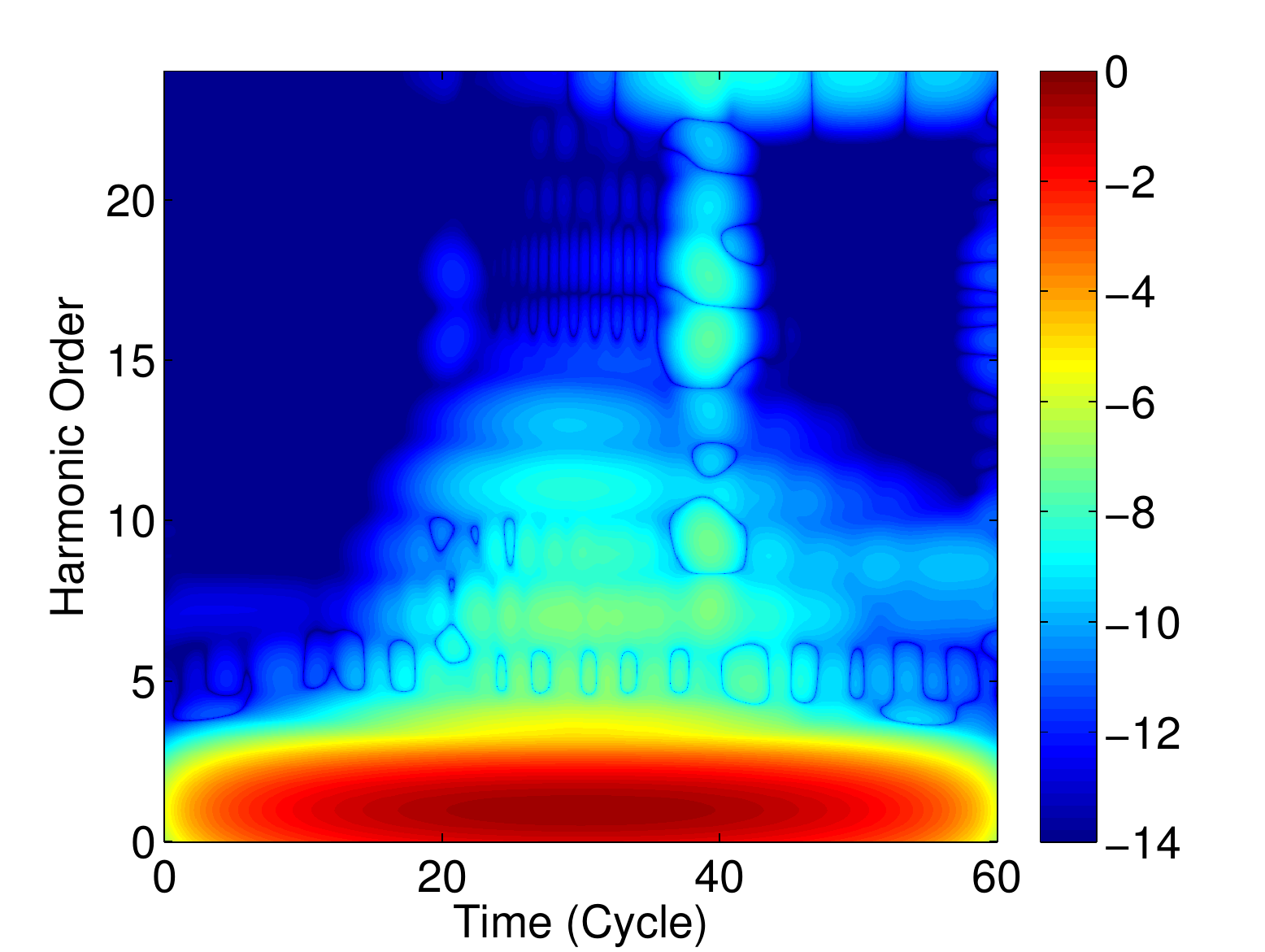}
		\hspace*{3.0cm}(c)\hspace{6.5cm}(d)\\
 \caption{ TF representations in the multiphoton ionization regime by the (a) GT, (b) MWT, (c) WVD, and (d) SPWVD. The lines indicating odd harmonics in the GT and MWT are subject to broadening issues arising from a window. Although the broadening artifact can be alleviated by the WVD, the evoked interference artifacts in the TF representation result in incomprehensible analysis. By applying additional filters, the SPWVD can moderate the interference patterns, at the price of broadening of the features in the TF representation.
 }
  \label{Fig2}
\end{figure*}

To unveil the dynamics of $d_L(t)$, first we apply the conventional linear and quadratic TF methods, as shown in Fig.~\ref{Fig2}.
The TF representation of the GT, computed by \eqn{intro.03} with $\sigma=57.94$ a.u., and the MWT, computed by \eqn{intro.03} with $\tau=30$, are displayed in Fig.~\ref{Fig2}(a) and Fig.~\ref{Fig2}(b), respectively.
Both the GT and MWT depict separate broad lines regarding the odd harmonics in the HHG process. 
While frequencies of the extracted components are consistent with the information shown in the Fourier spectrum in Fig.~\ref{Fig1}(d), the GT and MWT further capture the momentary behavior of each component.
For a fixed resolution in the GT, low frequency components could be lost given an insufficient window width.
In the TF representation of MWT, the same components are expected to be captured but the representation should be different from the GT due to the dilation nature of the transform; that is, the adaptive frequency resolution of the transform.
Indeed, the frequency resolution below the 7th harmonic is improved compared with Fig.~\ref{Fig2}(b), while harmonics on the upper TF plane remain broaden. In addition, the frequency-dependent weighting in $\sqrt\omega$ enhances the representation for high order harmonics.
Note the width of the window function $g(t)$ is uniform in the GT, and increases as $\omega$ increases in the MWT, as shown by the vertical lines between neighboring harmonics in the upper TF plane. 
The crescent shape distribution near the boundaries of the TF plane is caused by the boundary effect in CWT {-- for} a low frequency, the corresponding {scale} is {large} and the influence of the window cut-off {near the boundary} becomes apparent.
{The boundary effect is also inevitable in the GT, like the vertical artificial pattern in the very beginning and end. However it is less dominant.}

{As discussed above, while the window is avoided in the WVD, a strong interference pattern is inevitable} \cite{Book_TF_analysis}. 
The TF representation of the WVD, shown in Fig.~\ref{Fig2}(c), reveals components between the odd harmonics, which is inconsistent with the features in the Fourier spectrum and violates the parity symmetry in physics.
Moreover, harmonics above 15, of which intensities are weak, are not observed within the range of colorbar.
Interference along the time-direction is also observed.
As there are more {oscillatory} components and more overlaps {caused by} $x(s+\zeta/2){x^{*}(s-\zeta/2)}$ in \eqn{intro.06} as time increased, interference becomes stronger, as shown on the right-half TF plane in Fig.~\ref{Fig2}(c). 
Despite generating inaccurate information, note that in this simulation the WVD suggest the features of the TF representation, i.e., slow-varying components, regardless of a window parameter.
This suggest that the window width in the linear type TF methods should be large enough such that patterns depict the harmonics as that performed by the WVD.

In order to remove the interference, we employ the {SPWVD with} the filtering functions $g(s)=e^{\frac{-s^2}{2{\sigma_g}^2}}$ with $\sigma_g=100$ and $H(\xi)=e^{\frac{-\xi^2}{2{\sigma_H}^2}}$ with $\sigma_H=0.025$. Note that $g(0)=1$ and $H(0)=1$. 
The filtered result is given in Fig.~\ref{Fig1}(d). Although the even harmonics are eliminated, temporal interference cannot be fully removed. The introduced filters also worsen both temporal and frequency resolutions, as well as generating unexpected features. Note that the dynamic range (colorbar) is increased to reveal weak details.


\begin{figure*}
	  \includegraphics[width=0.495\hsize]{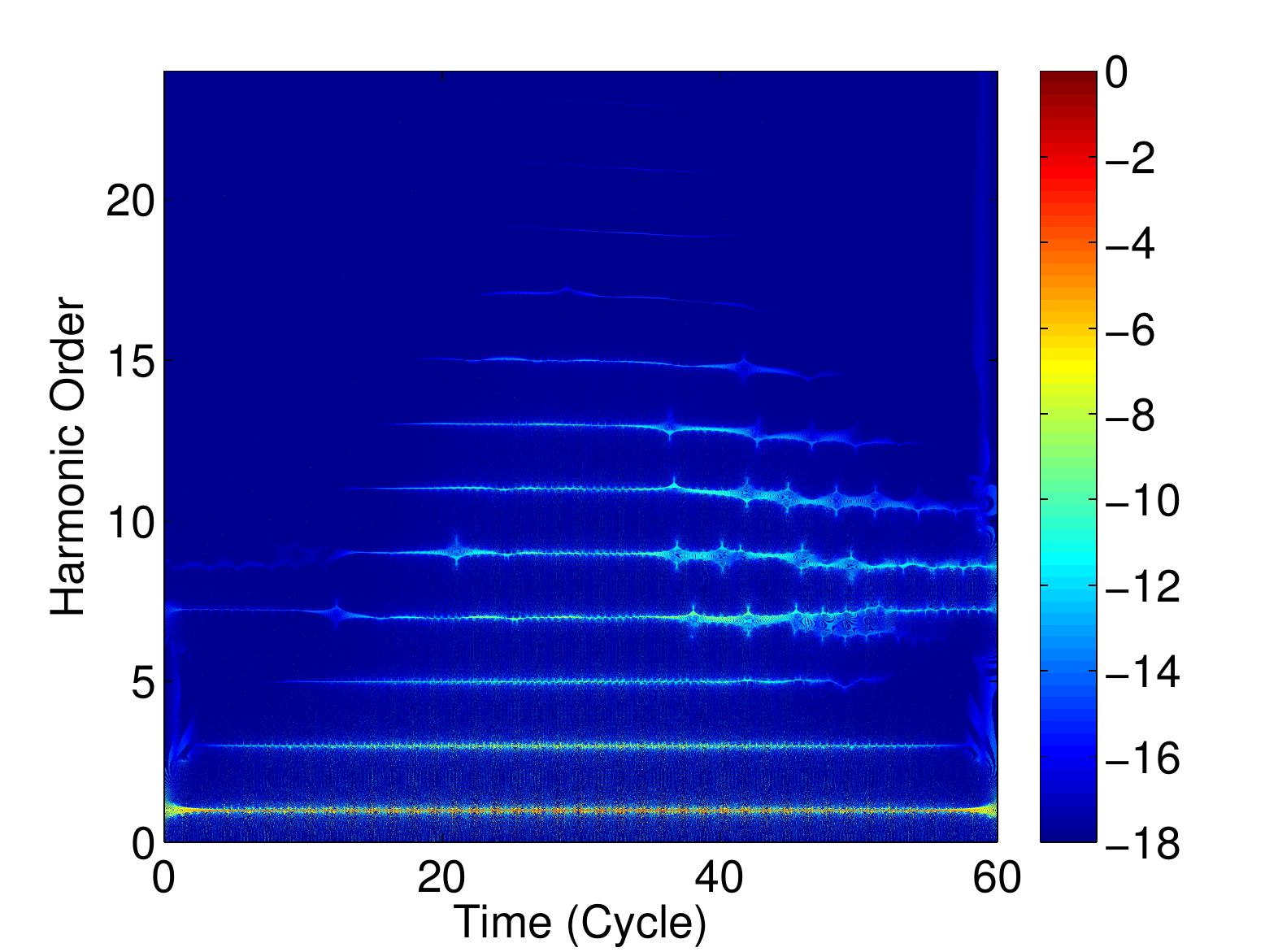}
		\includegraphics[width=0.495\hsize]{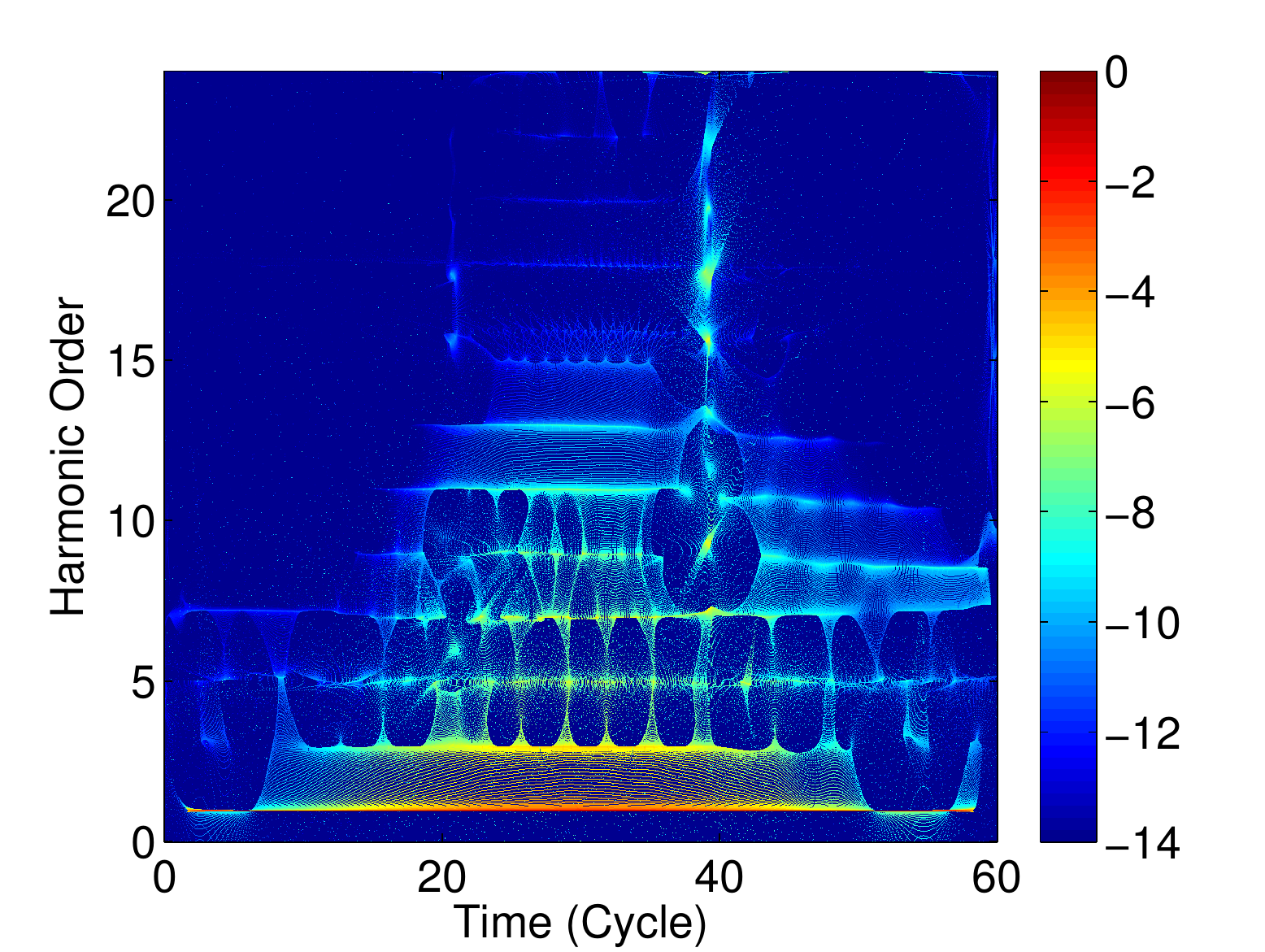}
		\hspace*{3.0cm}(a)\hspace{6.5cm}(b)\\
	  \includegraphics[width=0.495\hsize]{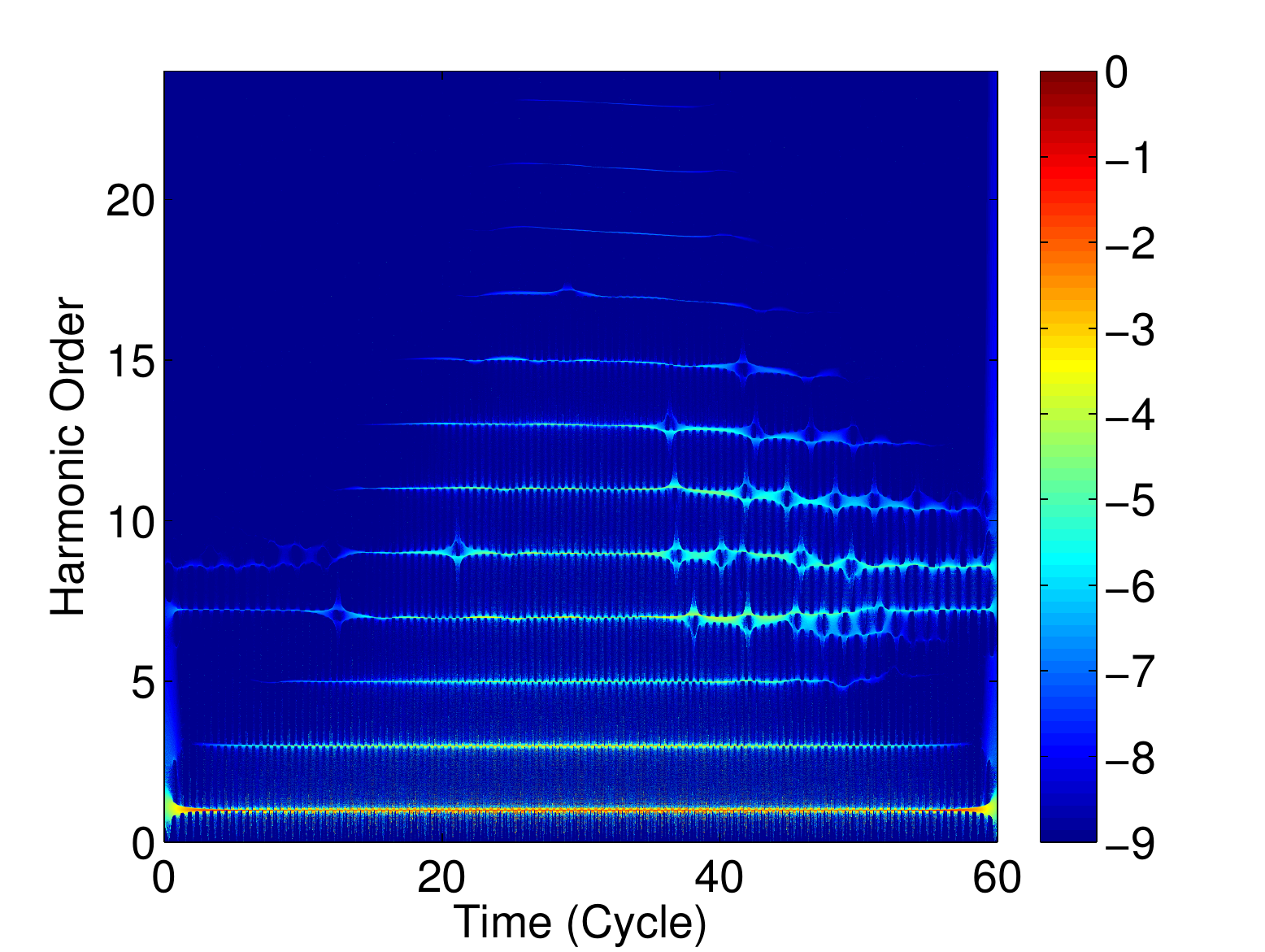}
		\includegraphics[width=0.495\hsize]{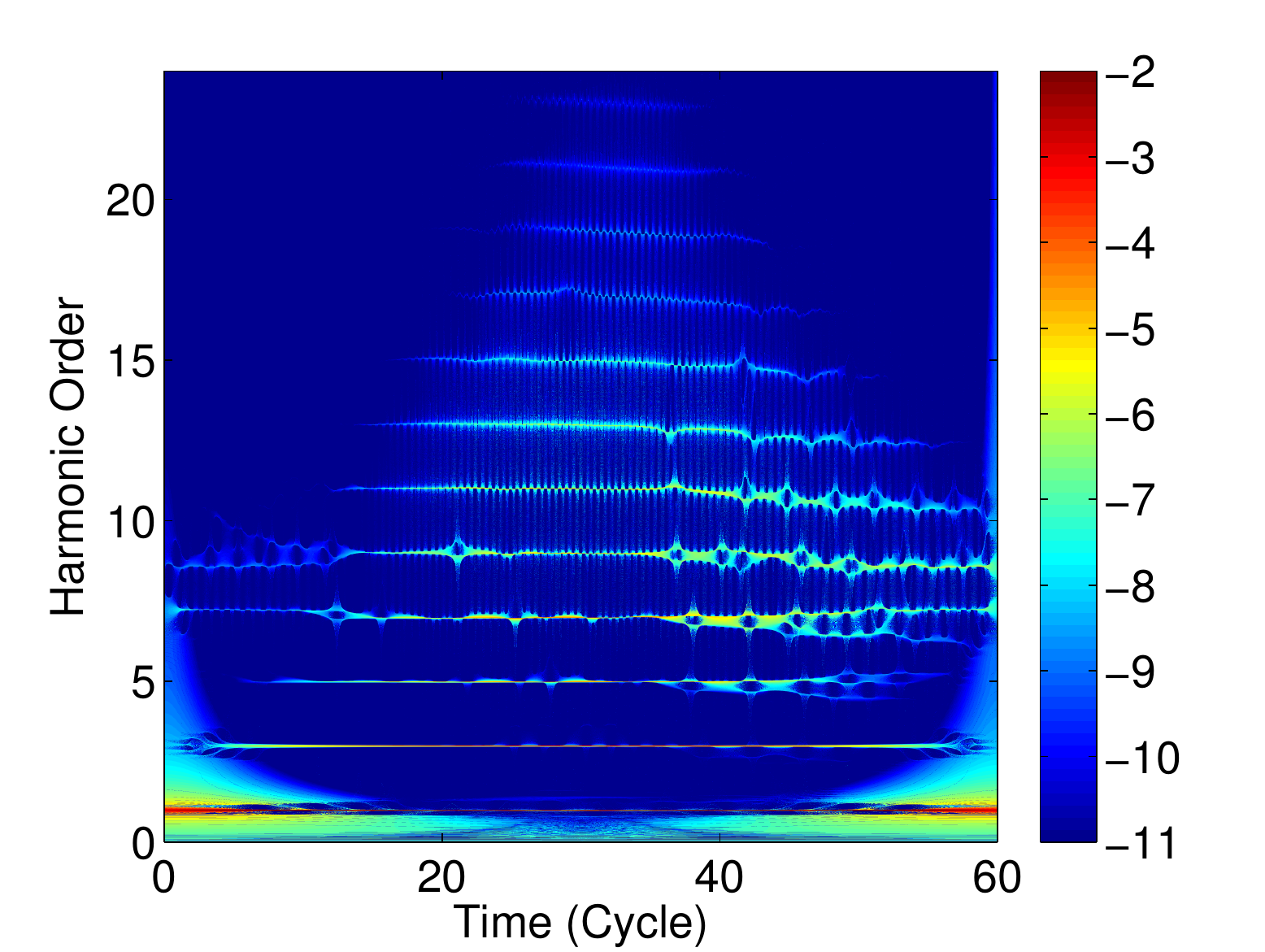}
		\hspace*{3.0cm}(c)\hspace{6.5cm}(d)\\
 \caption{ TF representations in the multiphoton ionization regime by the (a) RM-GT and (b) RM-SPWVD. The broadening caused by the window in the GT and filter functions in SPWVD is removed. Note that the frequency shift in the beginning $10$ cycles corresponds to the AC Stark effect. TF representations in the multiphoton ionization regime by the (c) SST-GT and (d) SST-MWT. The broadening issue caused by the window in GT and MWT are removed. Note that the frequency shift in the beginning $10$ cycles corresponds the AC Stark effect.
 }
  \label{Fig3}
\end{figure*}

By reassigning the energy distribution of the linear type TF methods and the SPWVD to their local centroids, the TF representations are sharpened. 
The results of reassigned GT (RM-GT) and reassigned SPWVD (RM-SPWVD) are shown in Fig.~\ref{Fig3}(a) and (b).
In Fig.~\ref{Fig3}(a), the TF representation of the RM-GT significantly improved the resolution, revealing clear and distinct odd harmonics, as well as their subtle variation.
Note that it is the square of the GT representation that is reassigned, therefore the range of the color bar differs from Fig.~{\ref{Fig2}(a)}.

In Fig.~\ref{Fig3}(b), although the broadening that comes with filtering functions are removed and the TF representation depicts odd harmonics similar to those in Fig.~{\ref{Fig3}(a)}, the interference artifact cannot be eradicated.
We further employ the SST on the GT (SST-GT) and MWT (SST-MWT), in which the allocation rule is a good approximation of the IF of each IMT. The results shown in Fig.~{\ref{Fig3}(c) and (d)} also depict similar features as that in the RM-GT.
Note that the SST-MWT preserves the adaptive resolution, weighted frequencies, and the boundary effect of the CWT.

In Fig.~\ref{Fig3}(a), (c) and (d), we show that the IMT functions can be decomposed without ambiguity by the RM-GT, SST-GT, and SST-MWT.
Although the RM can illustrate the distinct odd harmonics and frequency shift of the AC Stark effect by finding the local centroids, there is no routine for reconstructing the harmonics. 
SST, on the other hand, comes equipped with an inverse method (See \eqn{sst_STFT.05} and \eqn{sst_CWT.03}).

\begin{figure*}
	  \includegraphics[width=0.495\hsize]{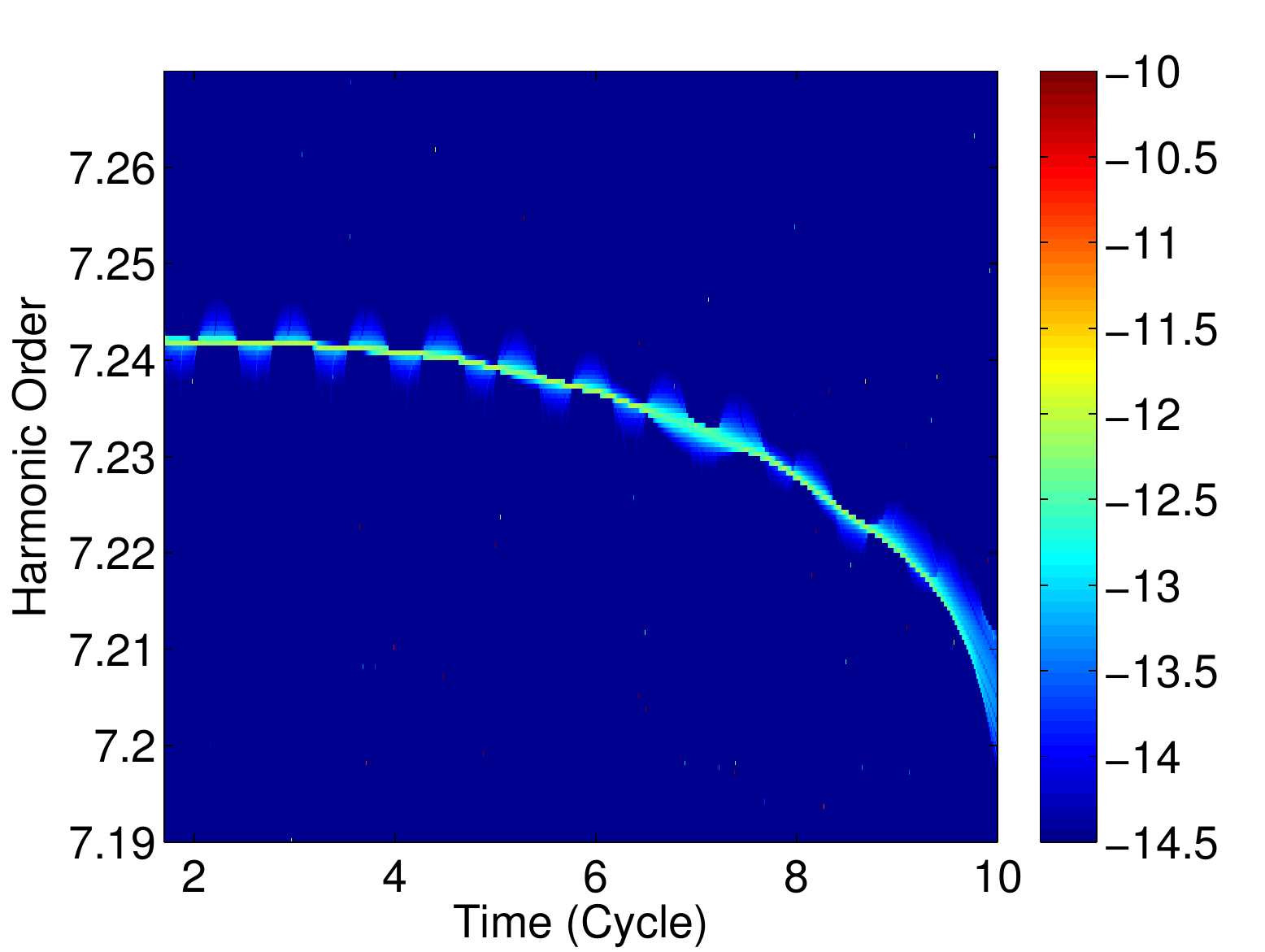}
		\includegraphics[width=0.495\hsize]{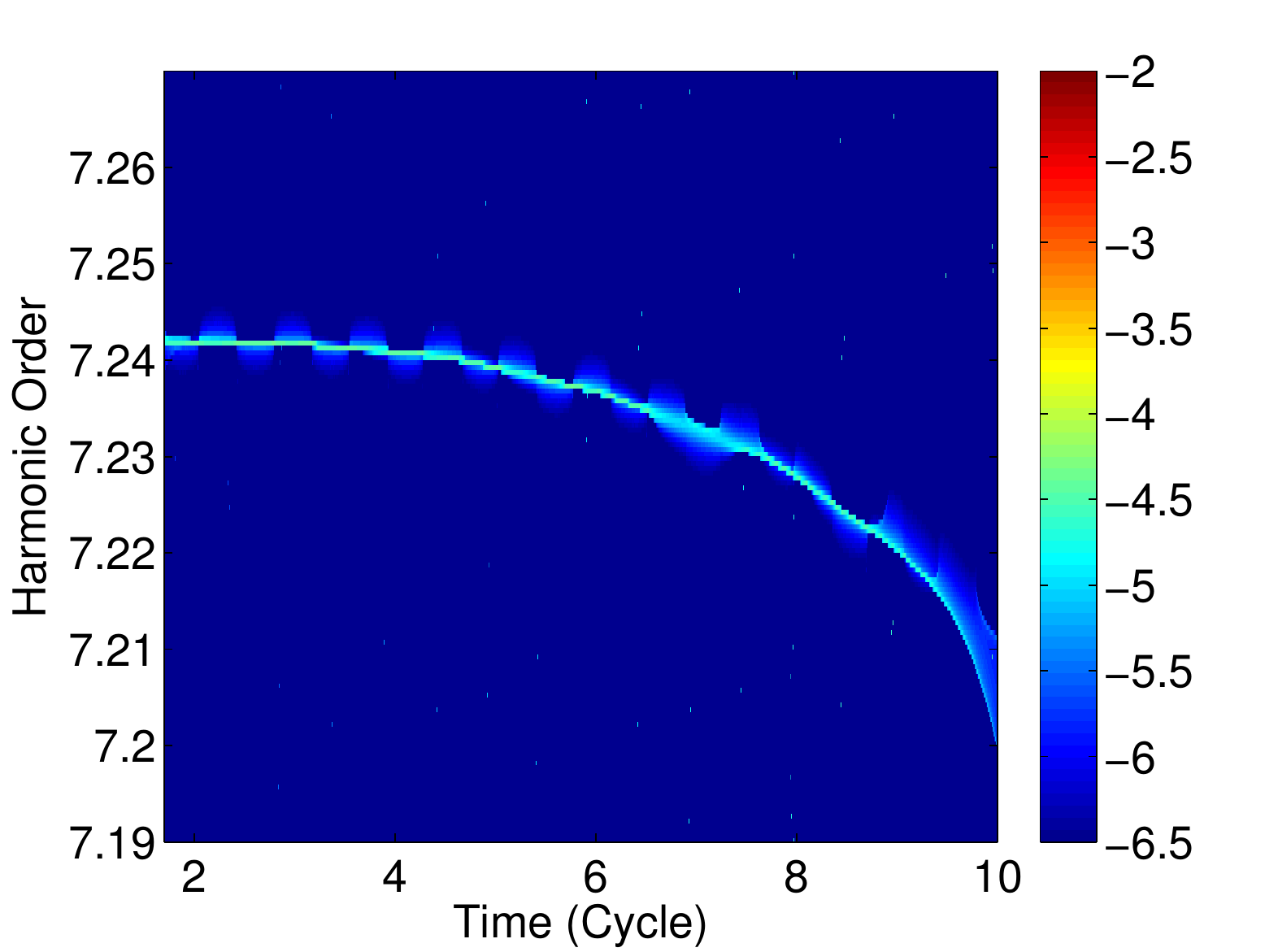}
		\hspace*{3.0cm}(a)\hspace{6.5cm}(b)\\
    \includegraphics[width=0.495\hsize]{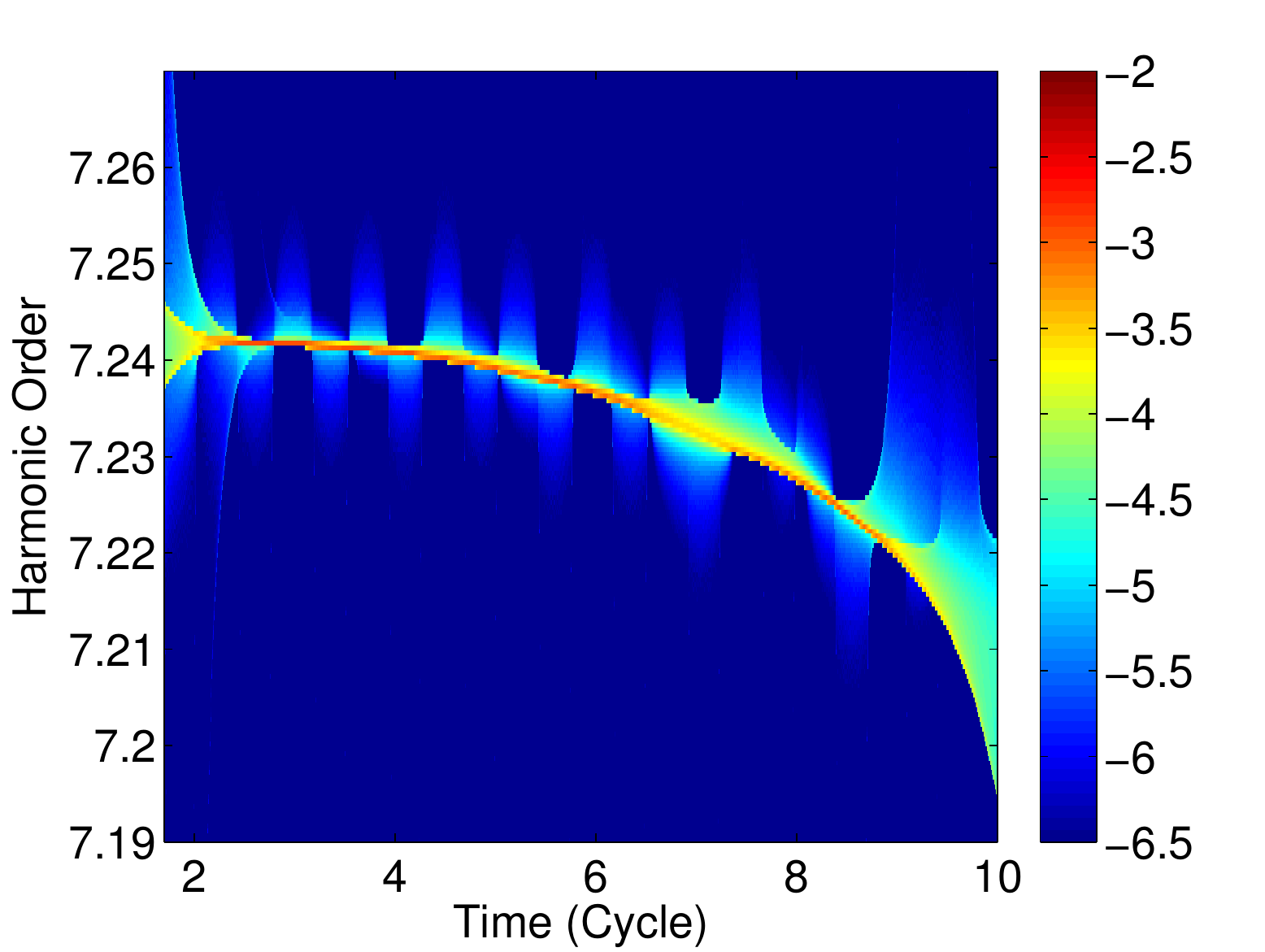}
		\includegraphics[width=0.495\hsize]{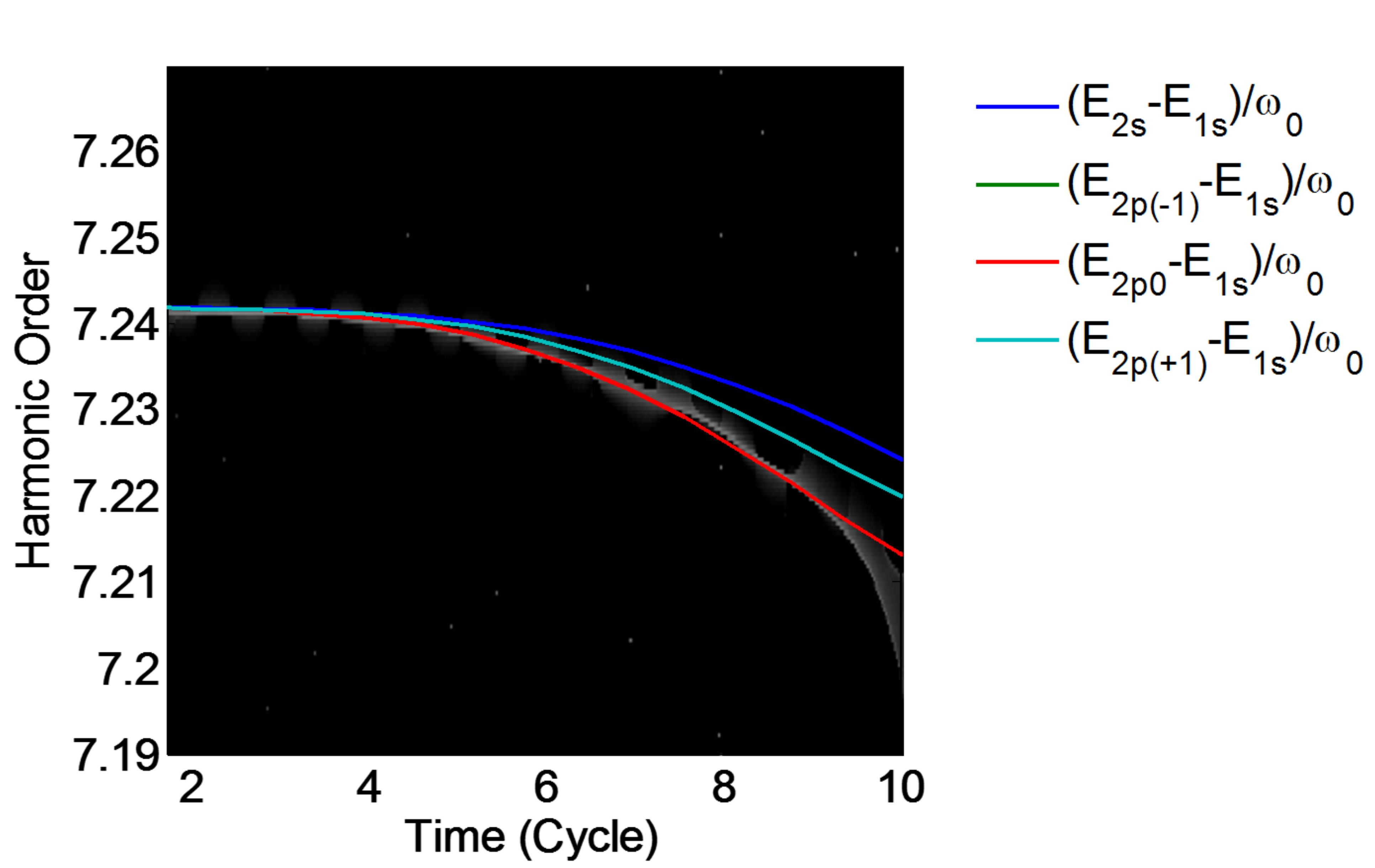}
		\hspace*{3.0cm}(c)\hspace{6.5cm}(d)\\
\caption{ The AC Stark effect revealed by the (a) RM-GT, (b) SST-GT and (c) SST-MWT. (d) Comparison of the frequency shift caused by the AC Stark effect computed by the SST-GT (gray scale) and the Floquet method. }
  \label{Fig5}
\end{figure*}

To explore the physical meaning of the shifting in Fig.~\ref{Fig5}(a)-(c), we analyze it using the Floquet method \cite{Shirley}, which has been extensively used in chemical physics \cite{Hsu, Hsu2}.
Details of the Floquet method can be found in Appendix B. 
In Fig.~\ref{Fig5}(d), the blue, green, red, and cyan lines denote the energy difference of 1s-2s, 1s-2$\rm{p}_{x}$, 1s-2$\rm{p}_{z}$, and 1s-2$\rm{p}_{y}$, respectively, computed by the Floquet method, in the unit of $\omega_0$.
Due to the breaking of the spherical symmetry by the laser field, the energy levels of 2s, 2$\rm{p}_x$, 2$\rm{p}_y$, and 2$\rm{p}_z$ shift and split. 
The quasi-energies derived by the Floquet computation and the curve by the SST-GT are superimposed for comparison. 
We found that the spectral line on the TF representations obtained by the SSTs not only demonstrate the AC Stark effect but also depict the selection rule, i.e., only the 1s-2$\rm{p}_{z}$ energy difference is presence.


It is worth mentioning that figures in Fig.~{\ref{Fig3}} illustrate a frequency shift at the beginning cycles of the 7th harmonic, which could be the AC Stark effect.
In the following context, we discuss about whether the IF components obtained by the reassigned values have a physical meaning, or they are simply clamped by artificial processing.

Details around the 7th harmonic are enlarged in Fig.~\ref{Fig5} for the (a) RM-GT, (b) SST-GT and (c) SST-MWT.
Note that the colorbar in  Fig.~\ref{Fig5}(a) is different from from that in  Fig.~\ref{Fig5}(b) and (c), since in the RM-GT the squared TF representation is reassigned.
Despite different intensities, the RM-GT, SST-GT and SST-MWT depict a similar shifting trend descending from the $7.241${-th} harmonic to the $7.200${-th} harmonic, which corresponds to the AC Stark effect.
The $7.241${-th} harmonic at the beginning cycles corresponds to the energies for 1s-2p transition ($\frac{1}{2}(1-\frac{1}{2^2})/\omega_0=7.241$ in a.u.), manifested as a small peak overlapping the H7 of the power spectrum in Fig.~\ref{Fig1}(d). 
The intensity of such spectral line is small because it arises from the near resonance absorption (not resonance absorption).


Fig.~\ref{Fig5}(b) and (c) suggest that the IFs shown by different SST methods are independent of {the chosen linear TF method}. 
Based on the independence of the chosen SST methods and the intimate matching of the analysis result and the theoretical prediction shown in Fig.~\ref{Fig5}(d), we conclude that the decomposed IMTs are physically meaningful.

\begin{figure}
\centering
	  \includegraphics[width=0.45\hsize]{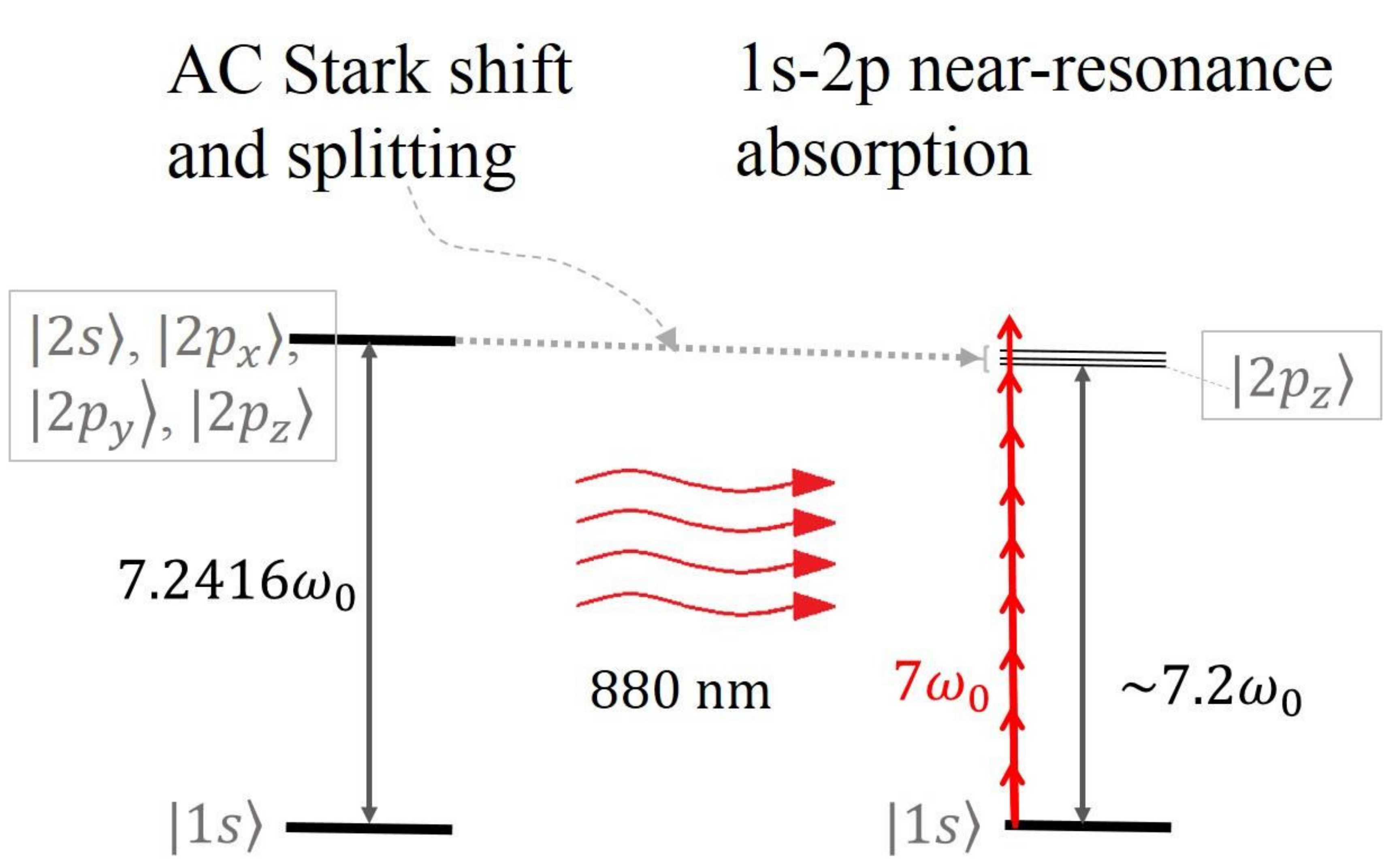}
	  \includegraphics[width=0.45\hsize]{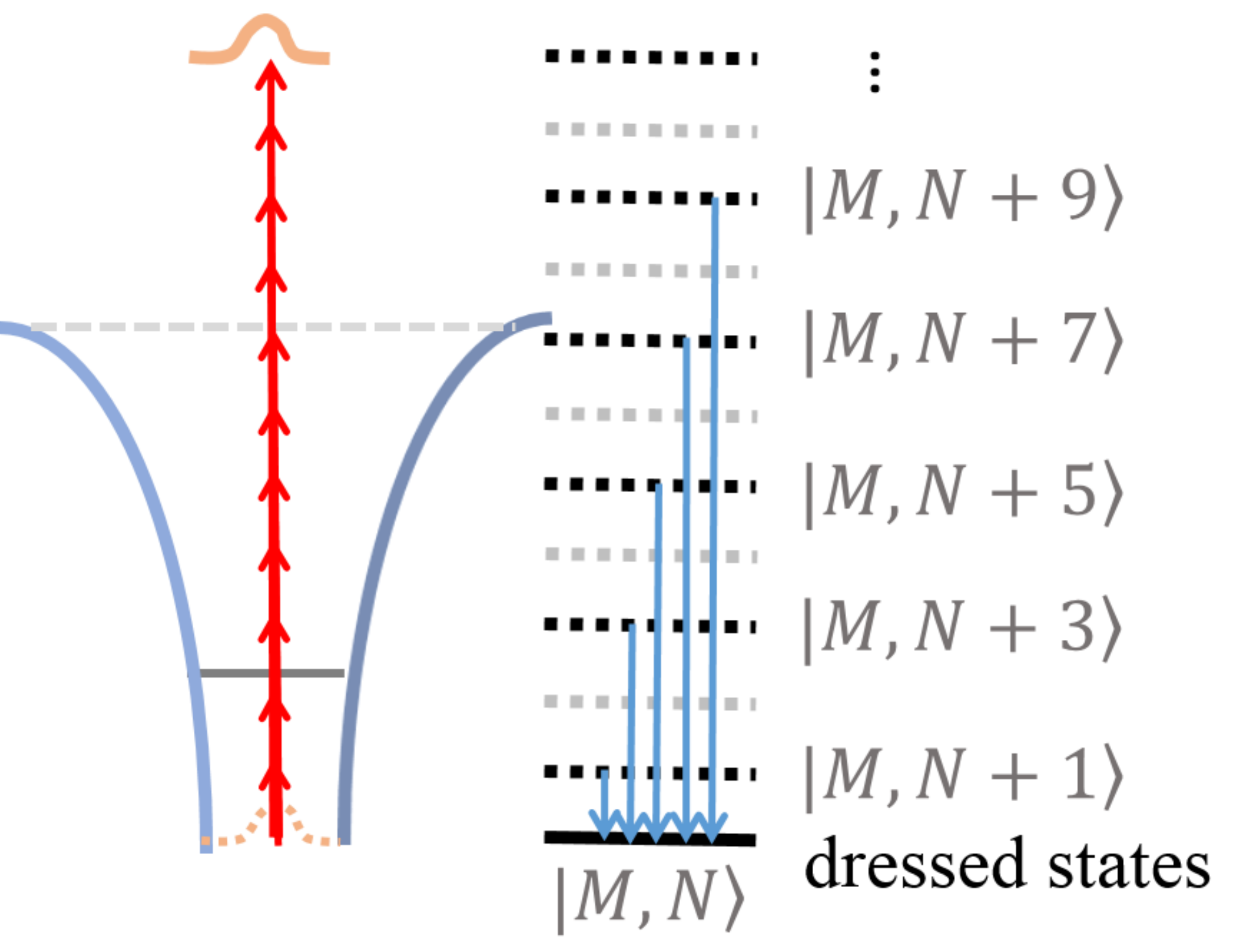}\\
		\hspace*{0.1cm}(a)\hspace*{4.2cm}(b)\\
\caption{ Illustration for the (a) AC-Stark effect and (b) the high-order harmonic process. }
  \label{multiphoton_ionization}
\end{figure}

The ionization process for the laser field in Fig.~\ref{Fig1}(a) is summarized in Fig.~\ref{multiphoton_ionization}.
In Fig.~\ref{multiphoton_ionization}(a), when the laser field is small, the spectral line $7.241$ aroused from the $1$s-$2$p near resonance absorption caused by the atomic structure.
As the laser intensity gradually increases, the energy levels of $2$s, $2\textrm{p}_\textrm{x}$, $2\textrm{p}_\textrm{y}$, and $2\textrm{p}_\textrm{z}$ shift and split due to the breaking of the symmetry by the electric field, which is regarded as the AC Stark effect.
When the laser intensity increases furthermore, the high-order harmonic process is induced and the dynamics is dominated by the transition between dressed states $|M,N\rangle$ formed by the electron state $M$ and the photon state $N$, as illustrated in Fig.~\ref{multiphoton_ionization}(b).


A summary for comparison of TF methods in the multiphoton ionization regime is provided in Table $2$.
Note that in this table we mention that the odd harmonics and the AC Stark effect cannot be obtained by EMD with the HS. In fact, we could not yet find physical meaning for the series of IMFs decomposed by the EMD in the multiphoton ionization regime.

\begin{table}
\label{list2}
\centering
\textbf{Table~2} Comparison of TF methods in the multiphoton ionization regime \\[1ex]
\begin{tabular}{lll}
Phenomenon& Discrete Odd Harmonics  & The AC Stark Effect  \\
\hline
GT      & Yes (blurred)       & Yes (blurred)     \\
MWT       & Yes (blurred)       & Yes (blurred)      \\
WVD       & No                  & No                 \\
SPWVD     & Yes, for               & Yes (blurred)  \\
          & harmonics $>5$ (blurred)  & \\
RM-GT   & Yes                 & Yes   \\
RM-SPWVD  & Yes                 & Yes (blurred)   \\
SST-GT  & Yes                 & Yes          	\\		
SST-MWT   & Yes                 & Yes          	\\	
EMD-HS    & No                  & No          \\
	
\end{tabular}
\begin{tablenotes}
\item []{In this study, the GT and MWT are examples for the STFT and CWT, respectively.   }
\end{tablenotes}
\end{table}


\subsection{Tunneling Ionization Regime}

\begin{figure*}
	  \includegraphics[width=0.495\hsize]{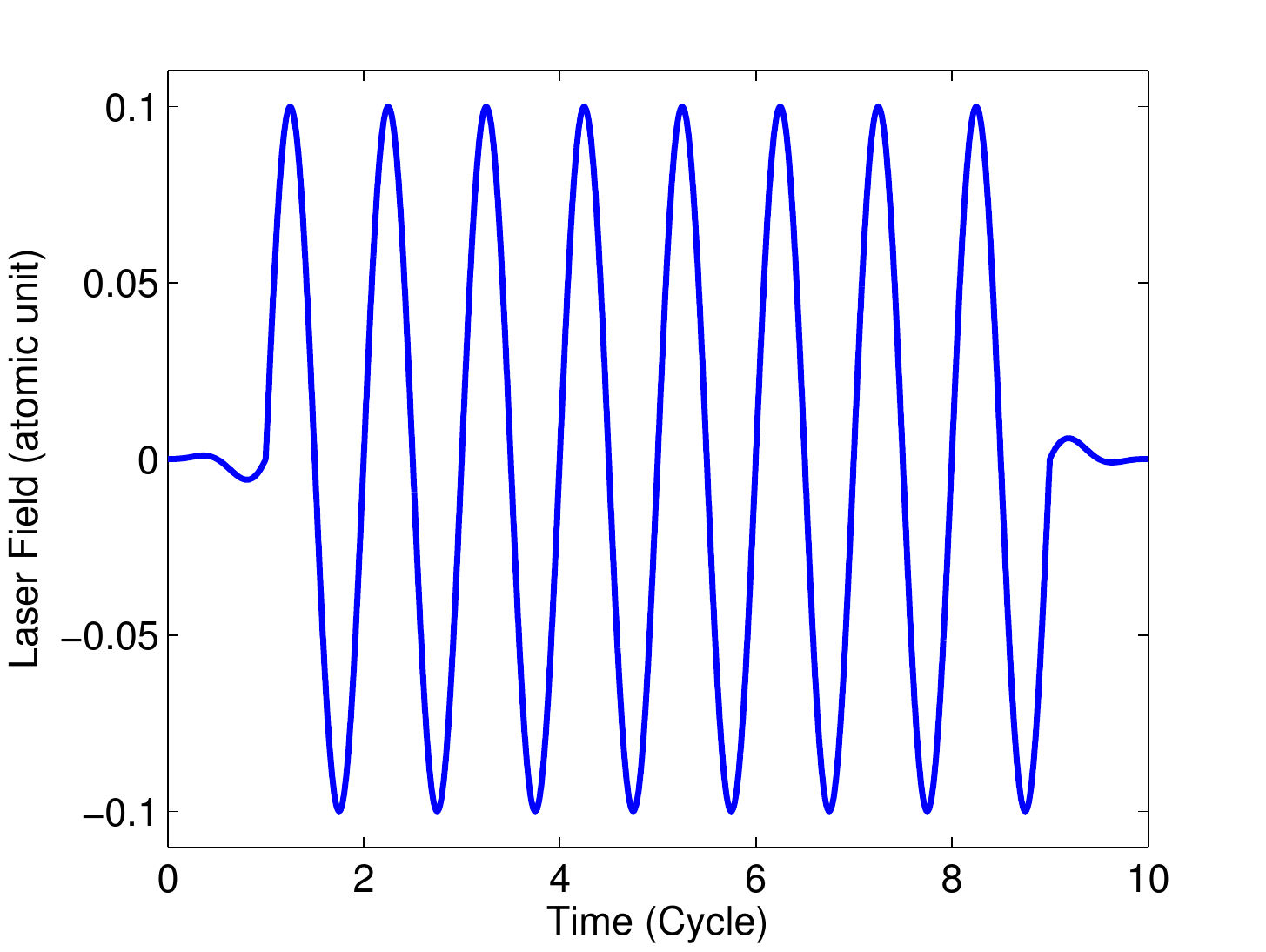}
		\includegraphics[width=0.495\hsize]{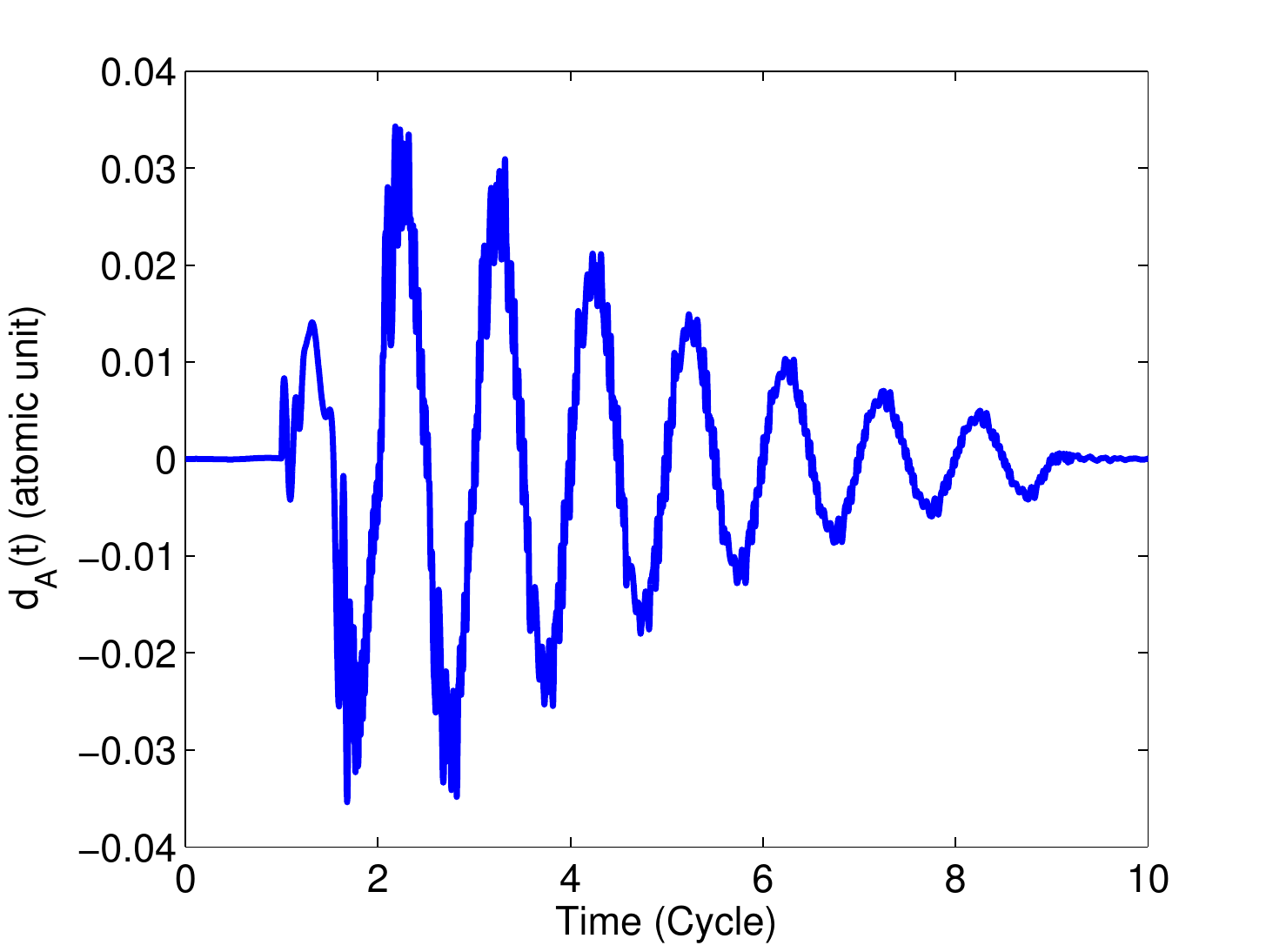}\\
				\hspace*{3.0cm}(a)\hspace{6.5cm}(b)\\
 \caption{ The simulation of laser-driven hydrogen in the tunneling ionization regime.  The laser wavelength is $800$ nm, and the laser intensity of $3.5\times10^{14}\;\mbox{W/cm}^2$, corresponding to the Keldysh parameter of $\gamma_K=0.57$. (a) The laser profile. (b) The induced dipole moment $d_A(t)$. 
 }
  \label{Fig1_Risoud}
\end{figure*}

In the second simulation, we discuss the TF representations for signals in the tunneling ionization regime.
The laser wavelength in this subsection is $\omega_0\approx0.05696100$ in atomic units (a.u.), corresponding to $800$ nm, and the laser intensity is $I_0=3.5\times10^{14}\;\mbox{W/cm}^2$. The Keldysh parameter is $\gamma_K=0.57$, suggesting that the tunneling ionization is dominant. The laser field has 10 cycles and its profile is ramped
on according to
\begin{eqnarray}
E_0(t) = \left\{\begin{array}{lll} \sin^2(\frac{\pi t}{10T})  &\mbox{when}\quad  0\leq t\leq T\\
       1  &\mbox{when}\quad  T\leq t \leq 9T\\
			\sin^2(\frac{\pi t}{10T})  &\mbox{when}\quad  9T\leq t\leq 10T\,,
\end{array}\right.	\label{Profile_Risoud}
\end {eqnarray}
which leads to the laser field $E_0F(t)\sin(\omega_0t)$, as shown in Fig.~\ref{Fig1_Risoud}(a).
We adopt the dipole moment in acceleration form in the case of tunneling ionization, as displayed in Fig.~\ref{Fig1_Risoud}(b).
The acceleration form can provide clearer information for the high order harmonics than the length form, because the differential operator in \eqn{eq.01a} acts as a high-pass filter.

\begin{figure*}
  	\includegraphics[width=0.495\hsize]{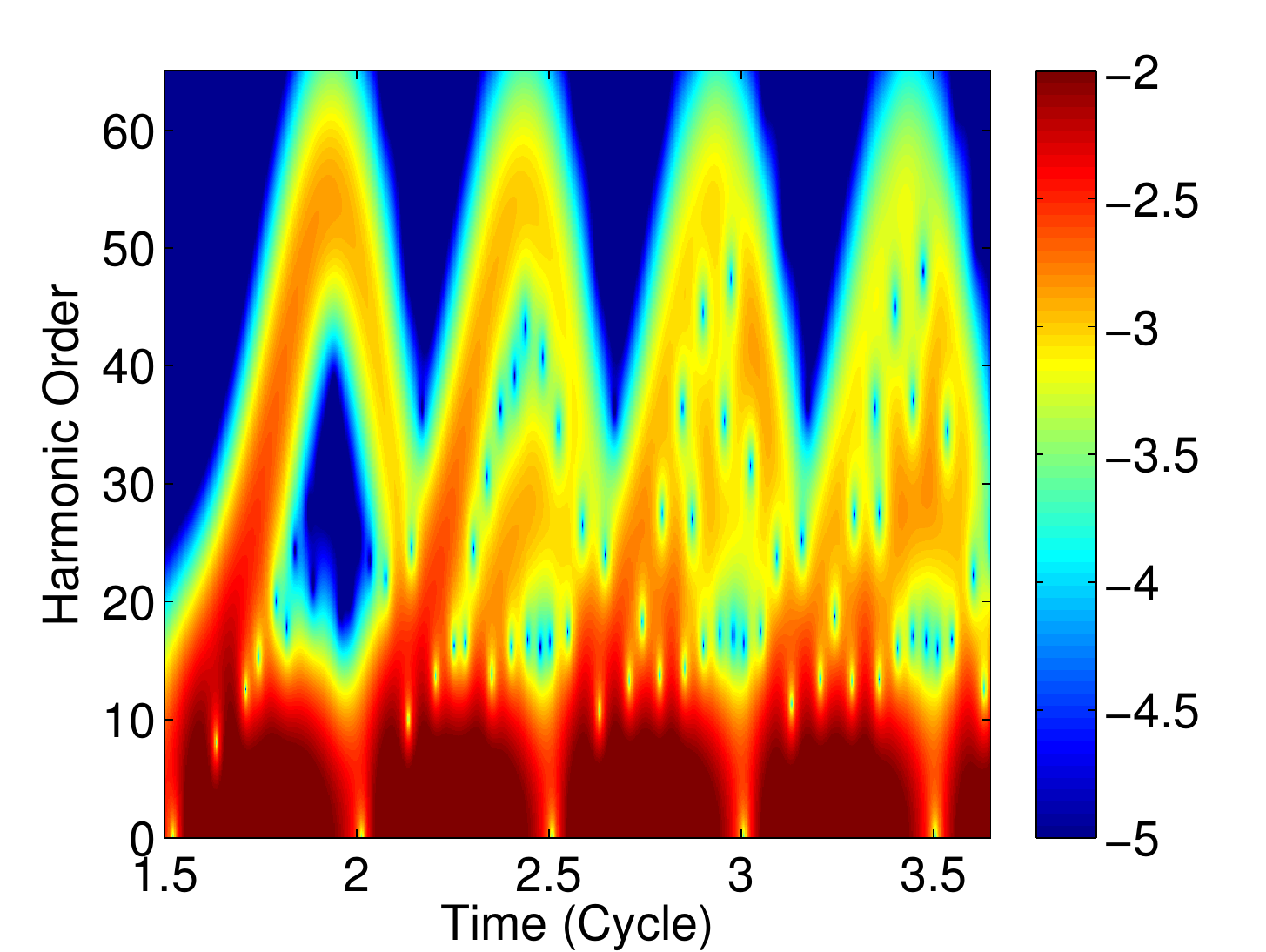}
	  \includegraphics[width=0.495\hsize]{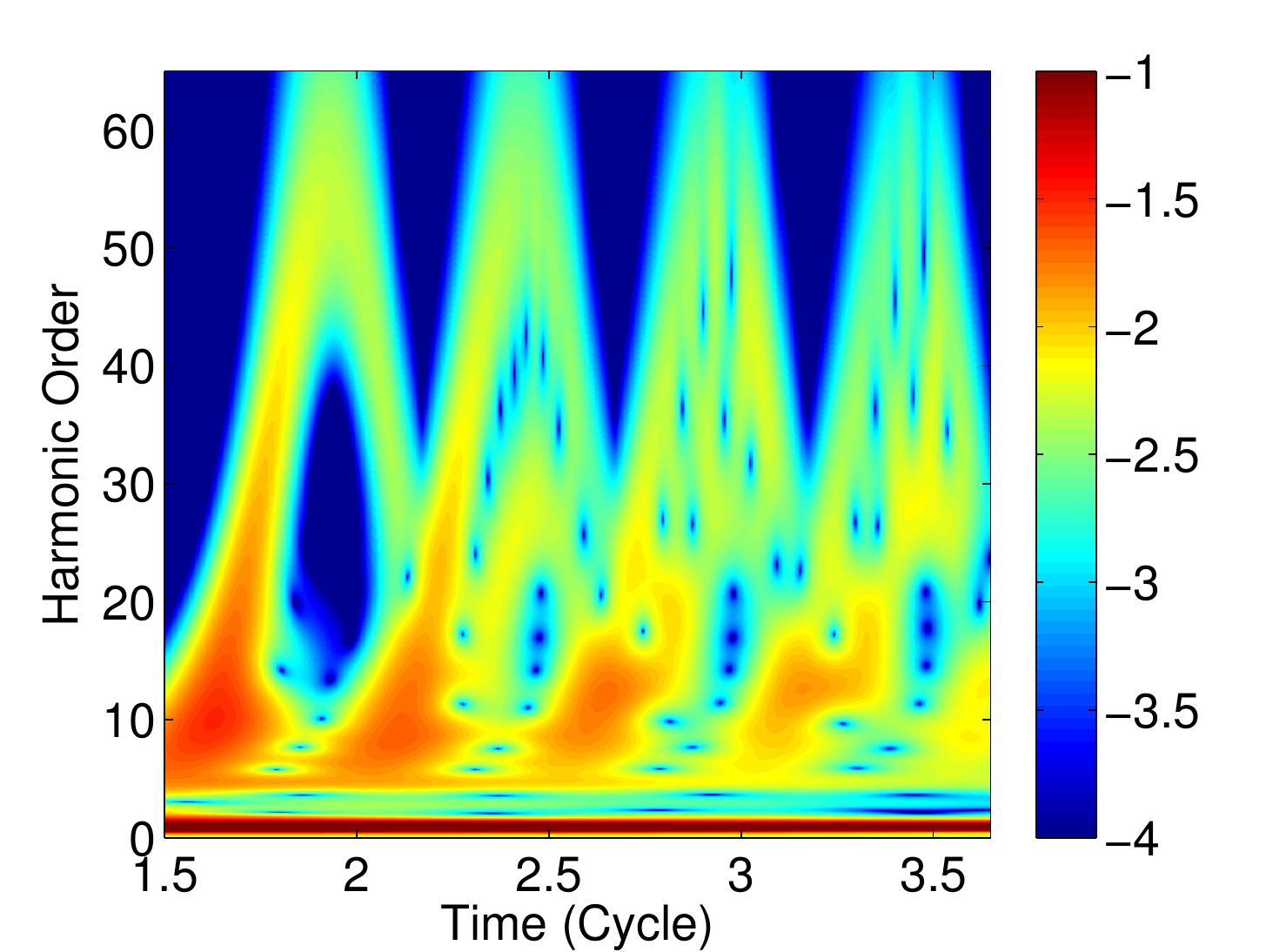}
		\hspace*{3.0cm}(a)\hspace{6.5cm}(b)\\
  	\includegraphics[width=0.495\hsize]{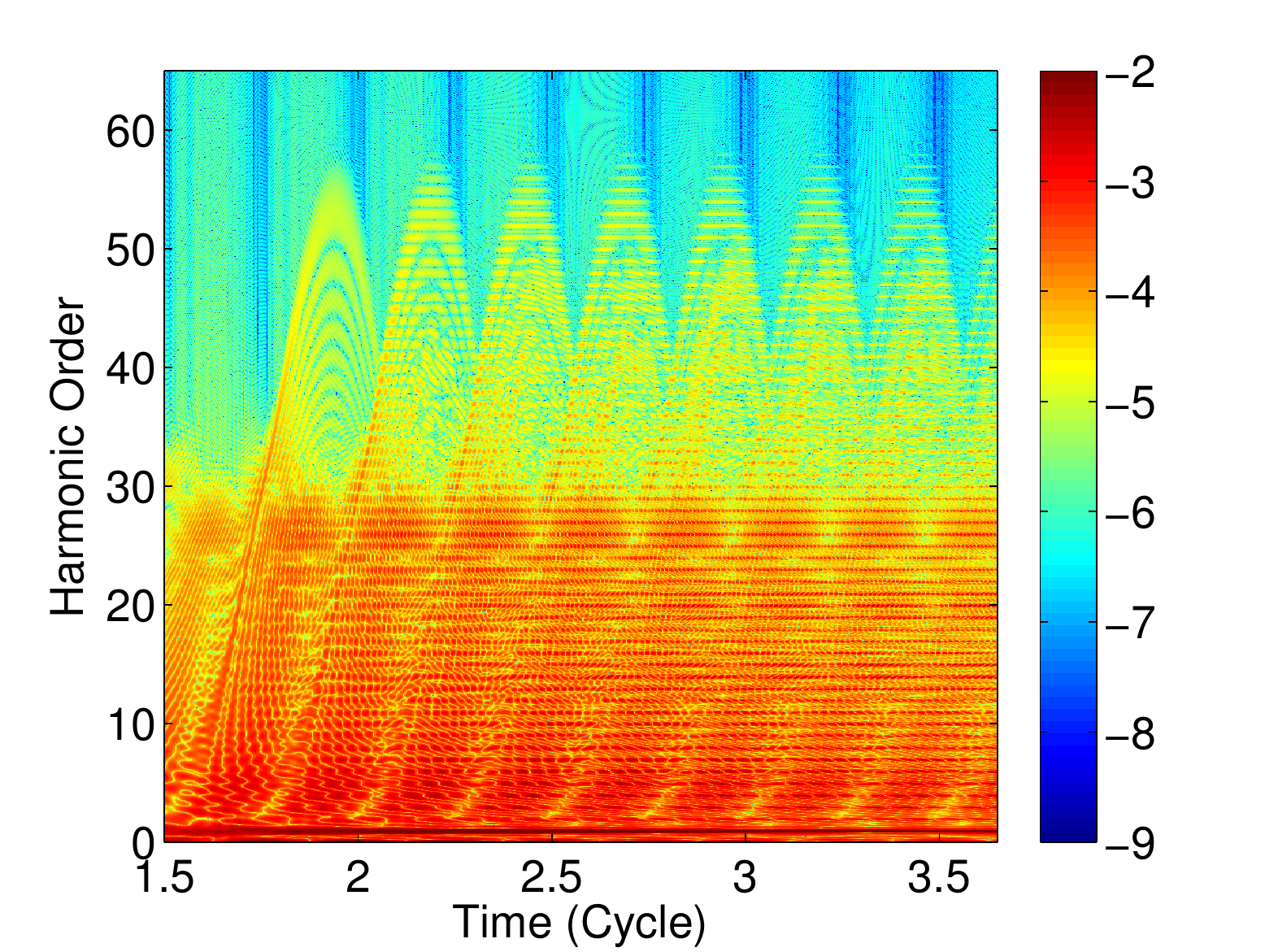}
	  \includegraphics[width=0.495\hsize]{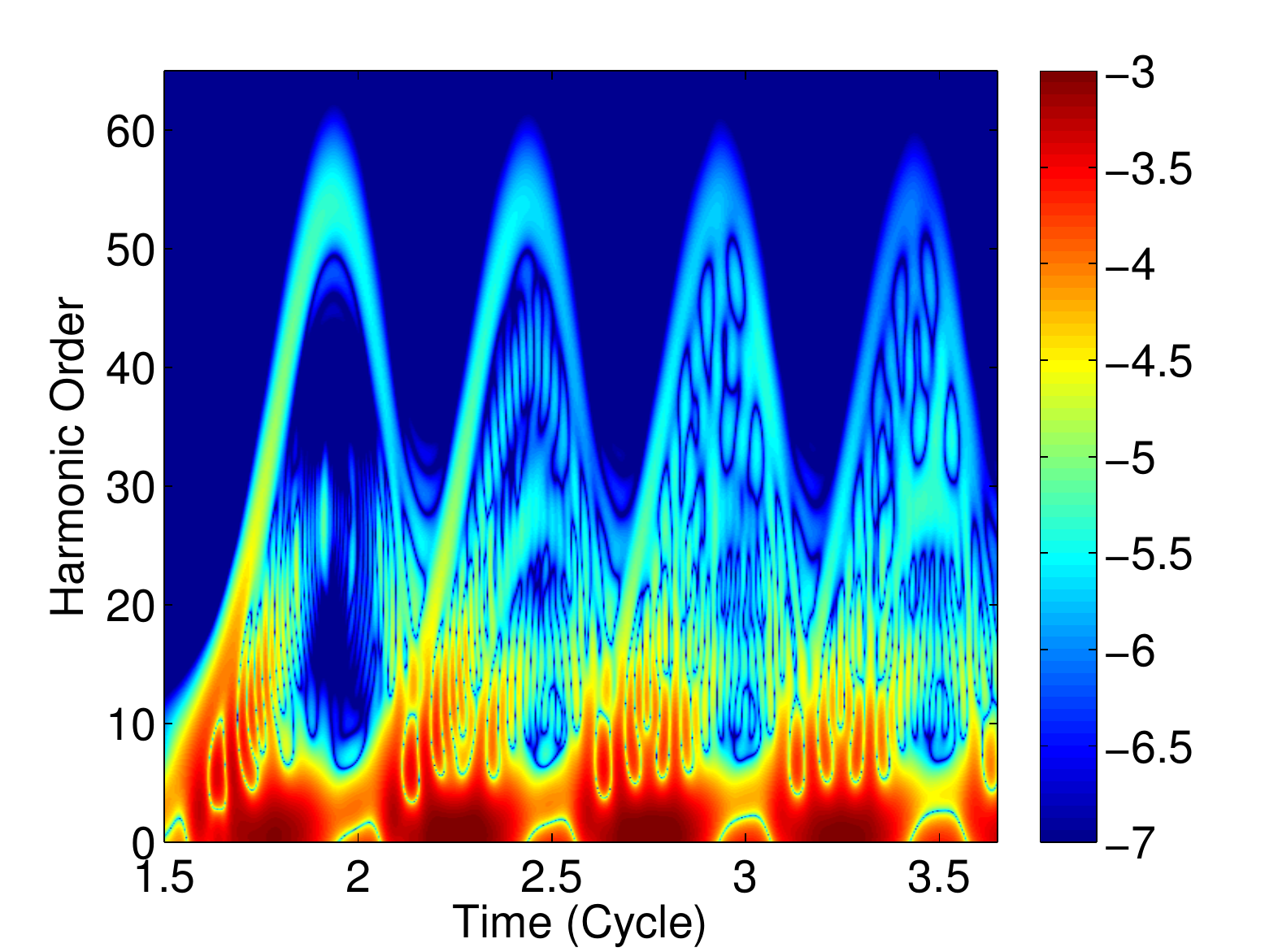}
  	\hspace*{3.0cm}(c)\hspace{6.5cm}(d)\\
 \caption{ TF representations in the tunneling ionization regime by the (a) GT, (b) MWT, (c) WVD, and (d) SPWVD. The repeating arches suggest excitation of the harmonics is strongly correlated with the laser field. In the representations of the GT and MWT, the arches are broaden by the window, leading to the difficulty to differentiate the intricate structures inside the arches. The representation of the WVD is obscured by the interference pattern, and it is difficult to remove the interference pattern with filters in the SPWVD.
 }
  \label{Fig6}
\end{figure*}

TF representations for the linear type and quadratic type of transforms are presented in Fig.~\ref{Fig6}, including the GT (Fig.~\ref{Fig6}(a)), MWT (Fig.~\ref{Fig6}(b)), WVD (Fig.~\ref{Fig6}(c)), and SPWVD (Fig.~\ref{Fig6}(d)).
In Fig.~\ref{Fig6}(c), we observe a periodic repetition of arches, suggesting the chirp-like dynamics of the attosecond radiation, as predicted by the standard semiclassical model \cite{Atto2}. 
To present such feature in the linear type TF representations, a short window width and a small quality factor $\tau$ are needed for the GT and the MWT, respectively. (Applicability of a short window width in the IMT models deserves further investigation.)
Here we select $\sigma=1$ (a.u.) for the GT and $\tau=6$ for the MWT.
TF representations in the GT and the MWT illustrate that the arch structure repeats every $0.5$T.
The inner structure after the 2nd cycle is indistinct because of the broadening caused by the window. 
While the GT cannot resolve information below the 15th harmonic due to the small window width, the MWT can describe the near and the below threshold harmonics (low order harmonics) by its merit of adaptive resolution.
WVD, on the other hand, can characterize the chirp very well without the need of window parameter and the broadening that comes along.
However, as described in previous subsection, WVD introduces strong interferences in both temporal and frequency directions.
Despite in SPWVD, where we use the parameters $\sigma_g=1$ and $\sigma_H=0.13$, it does not seem possible to eliminate the interference that mingled with intrinsic structure inside the main arches.


\begin{figure*}
	  \includegraphics[width=0.495\hsize]{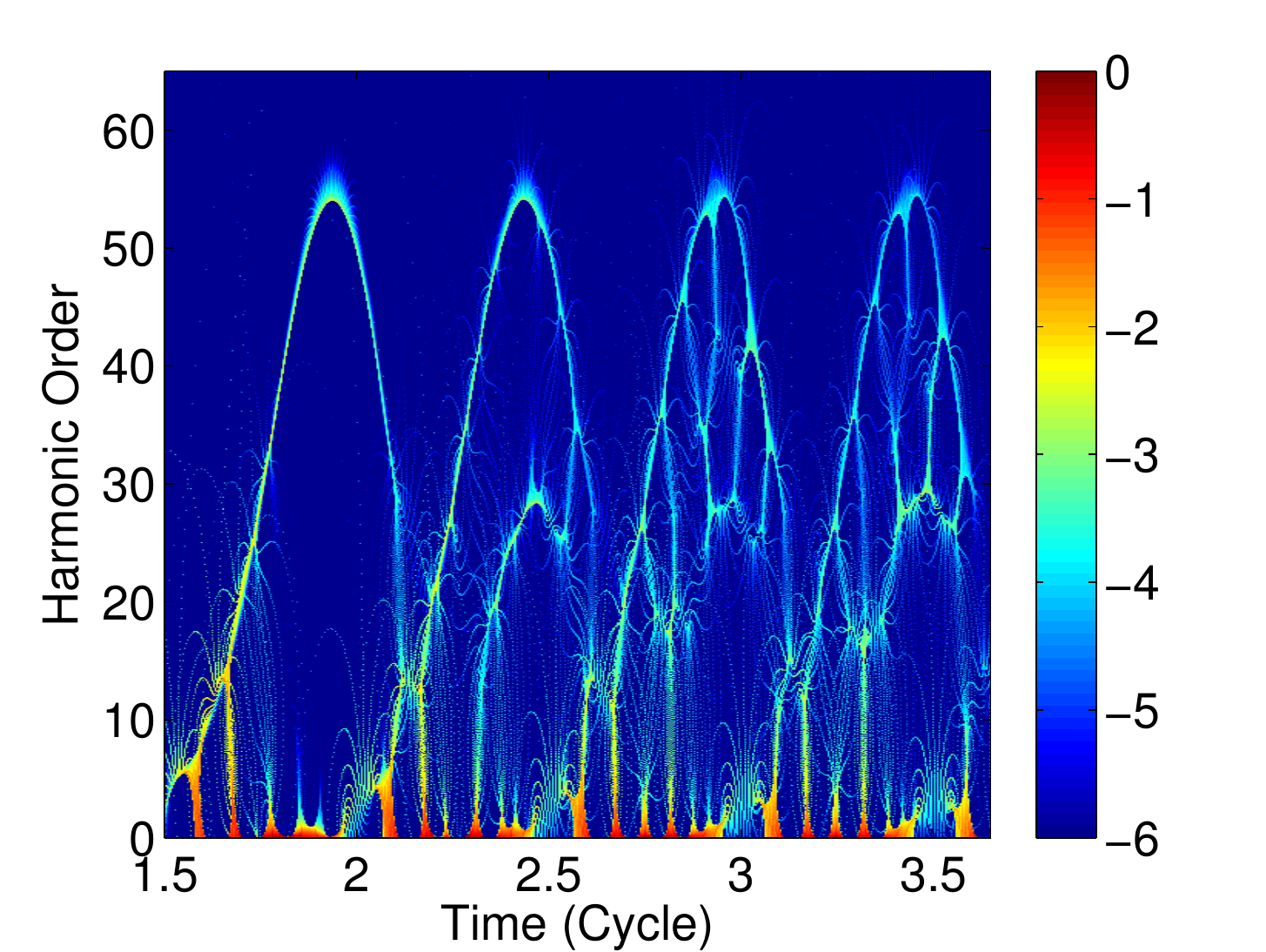}
		\includegraphics[width=0.495\hsize]{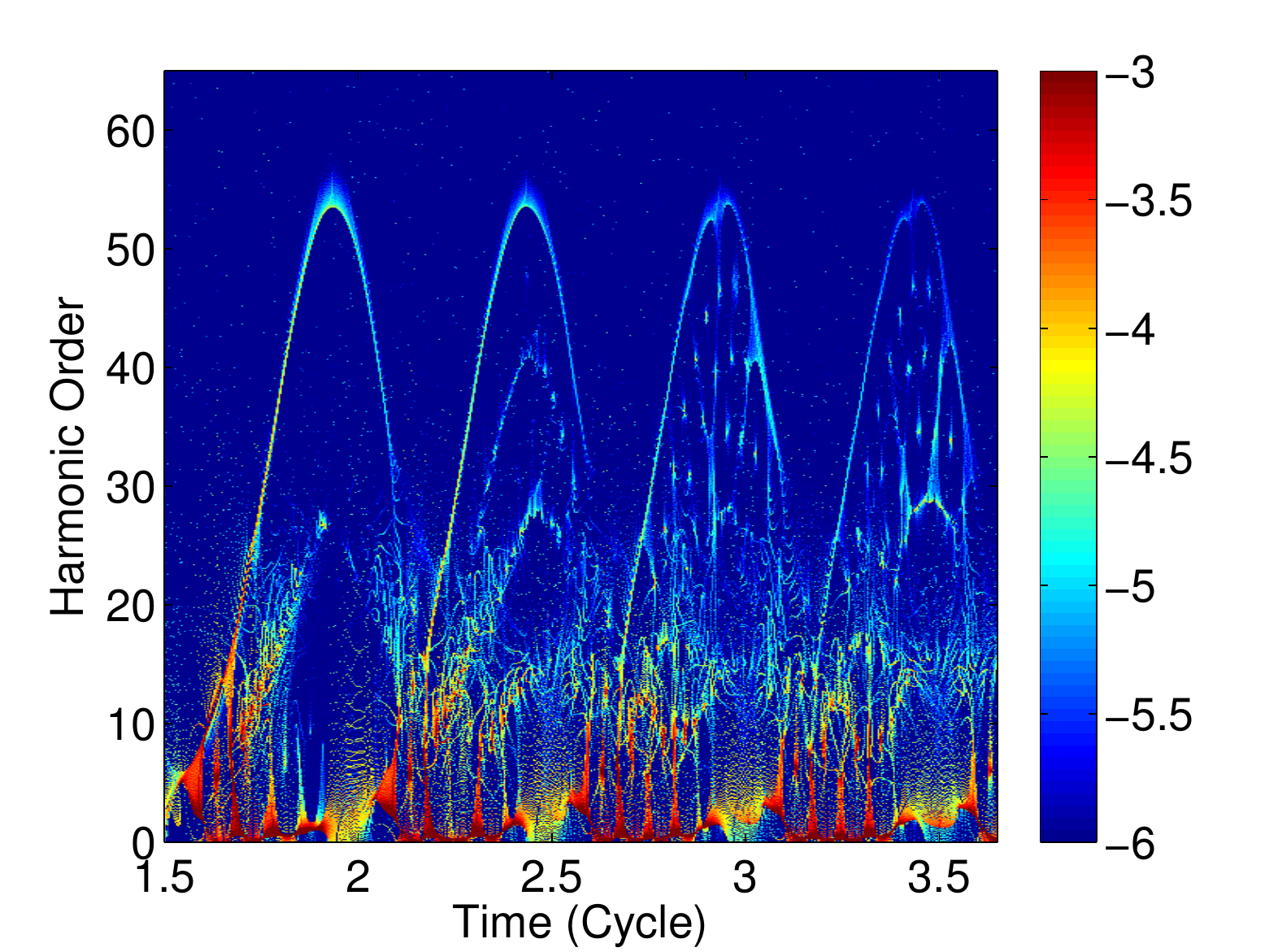}
  	\hspace*{3.0cm}(a)\hspace{6.5cm}(b)\\
  	\includegraphics[width=0.495\hsize]{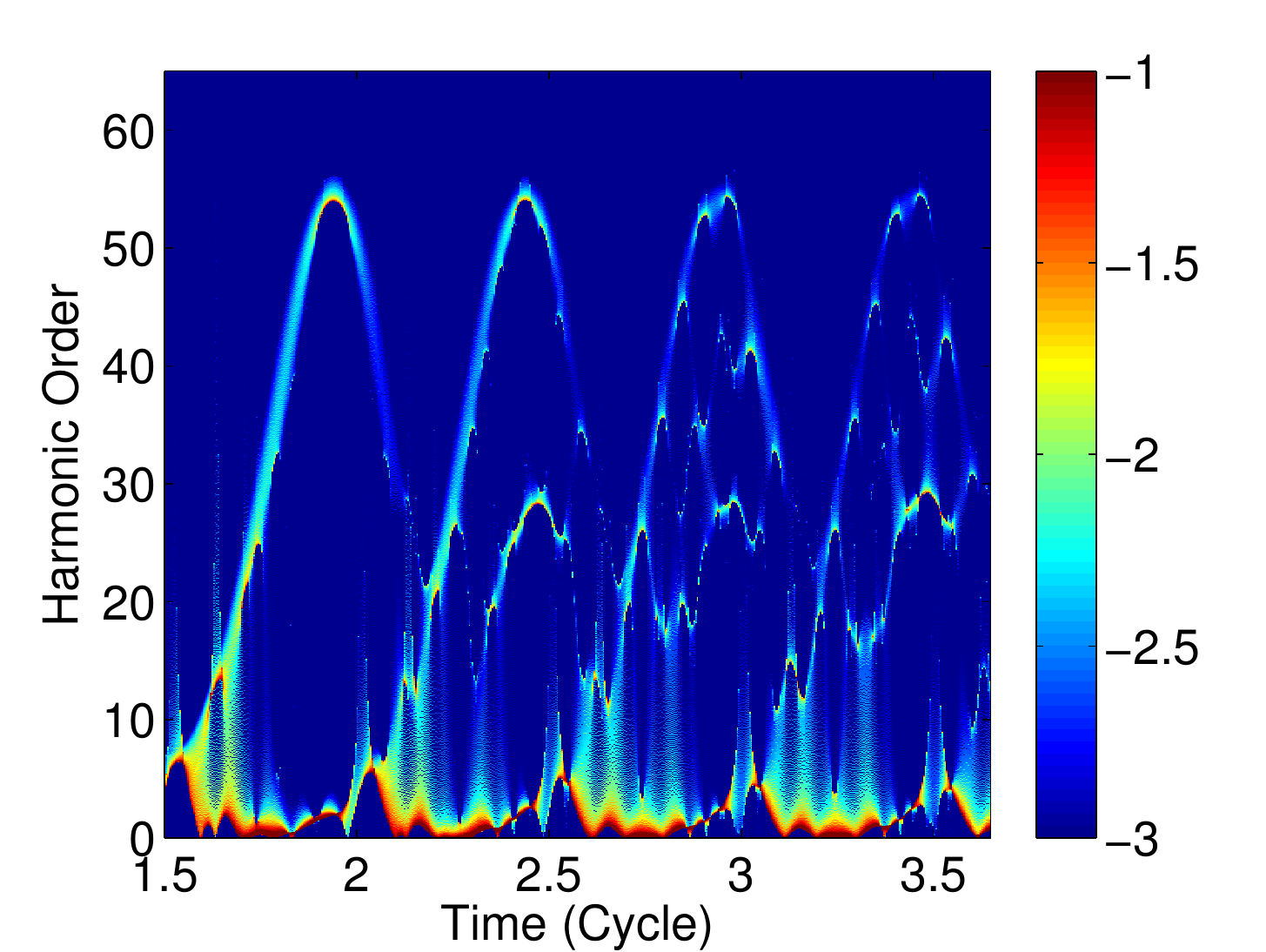}
	  \includegraphics[width=0.495\hsize]{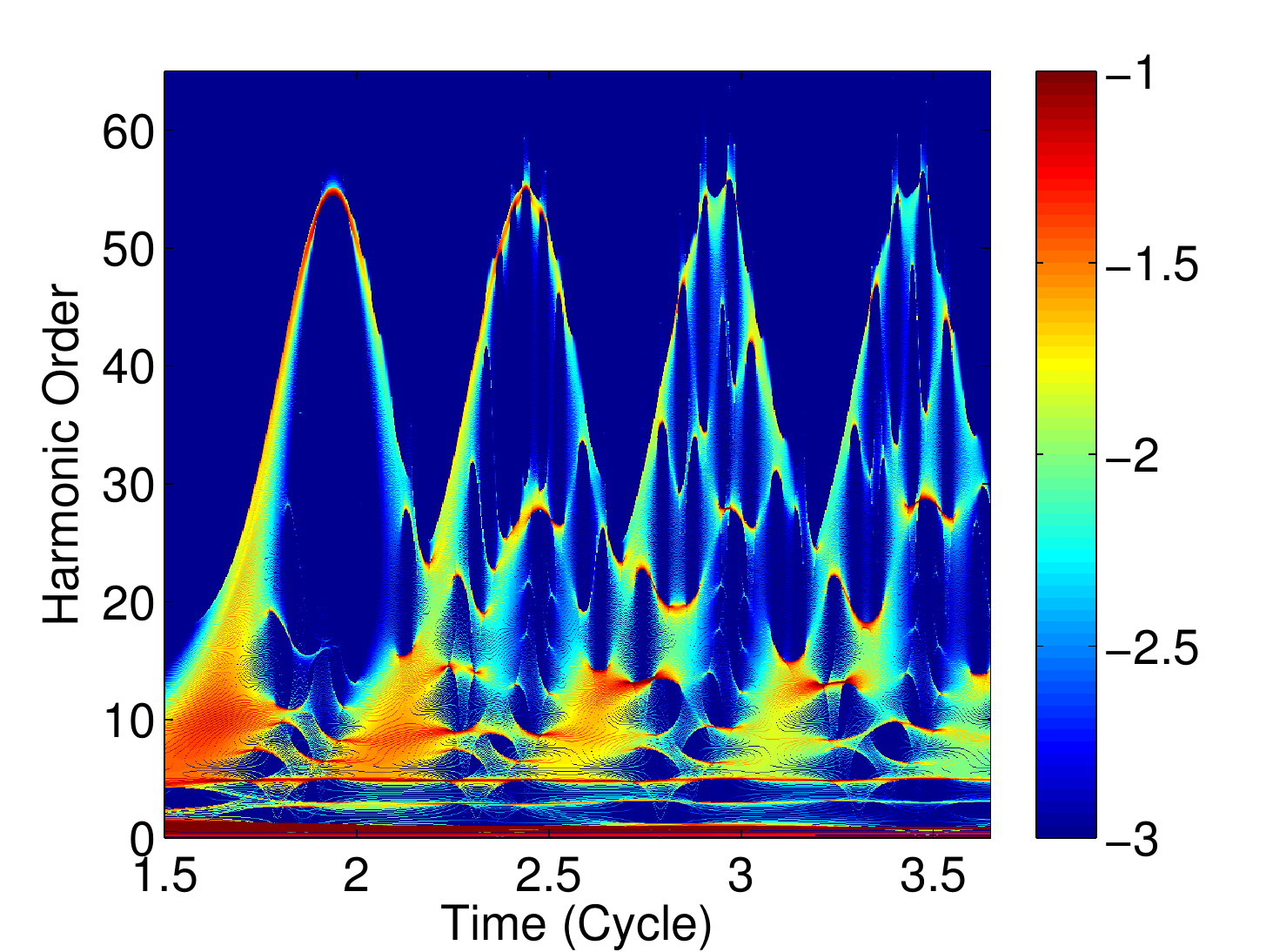}
  	\hspace*{3.0cm}(c)\hspace{6.5cm}(d)\\
  	\includegraphics[width=0.495\hsize]{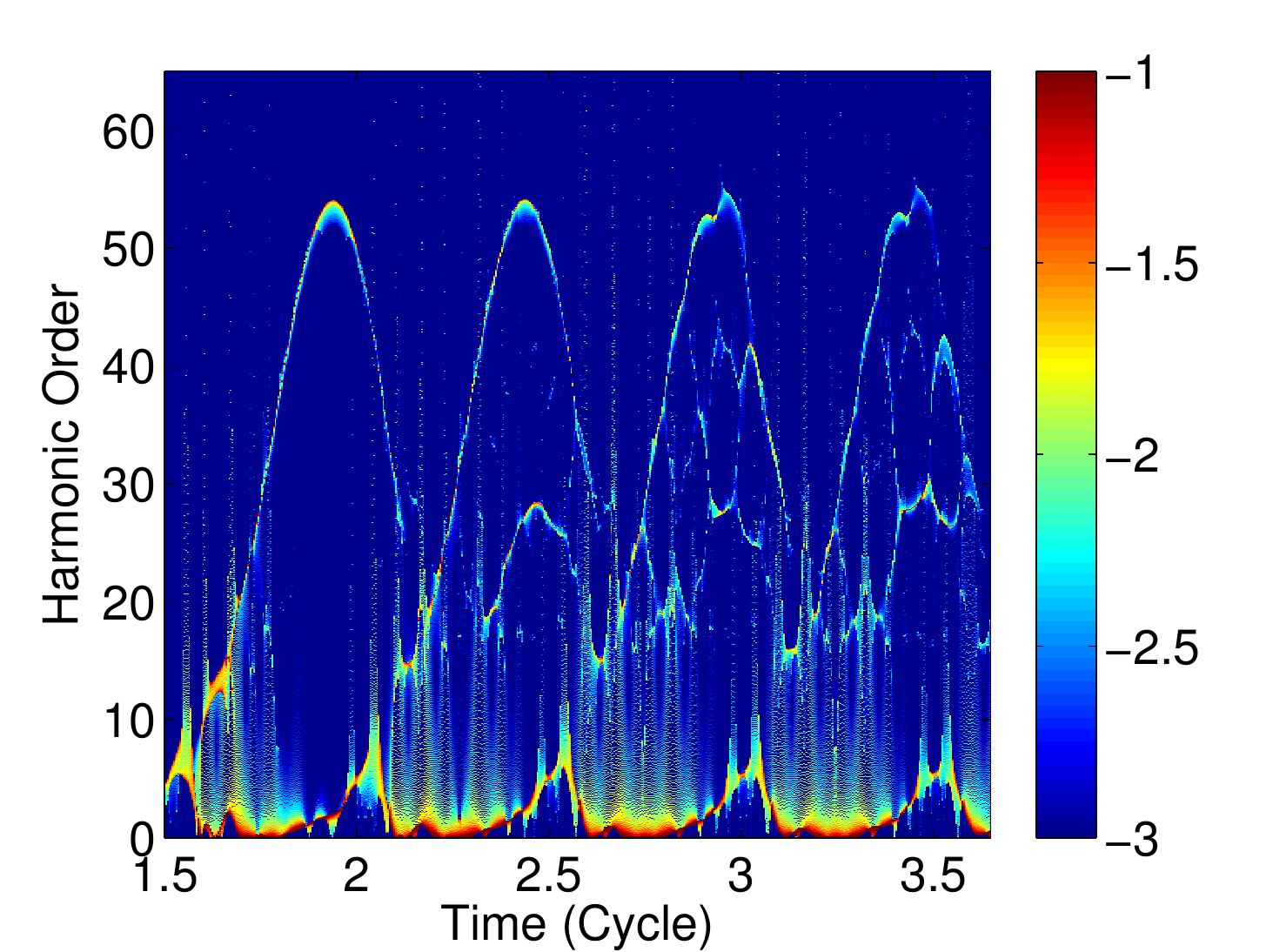}
	  \includegraphics[width=0.495\hsize]{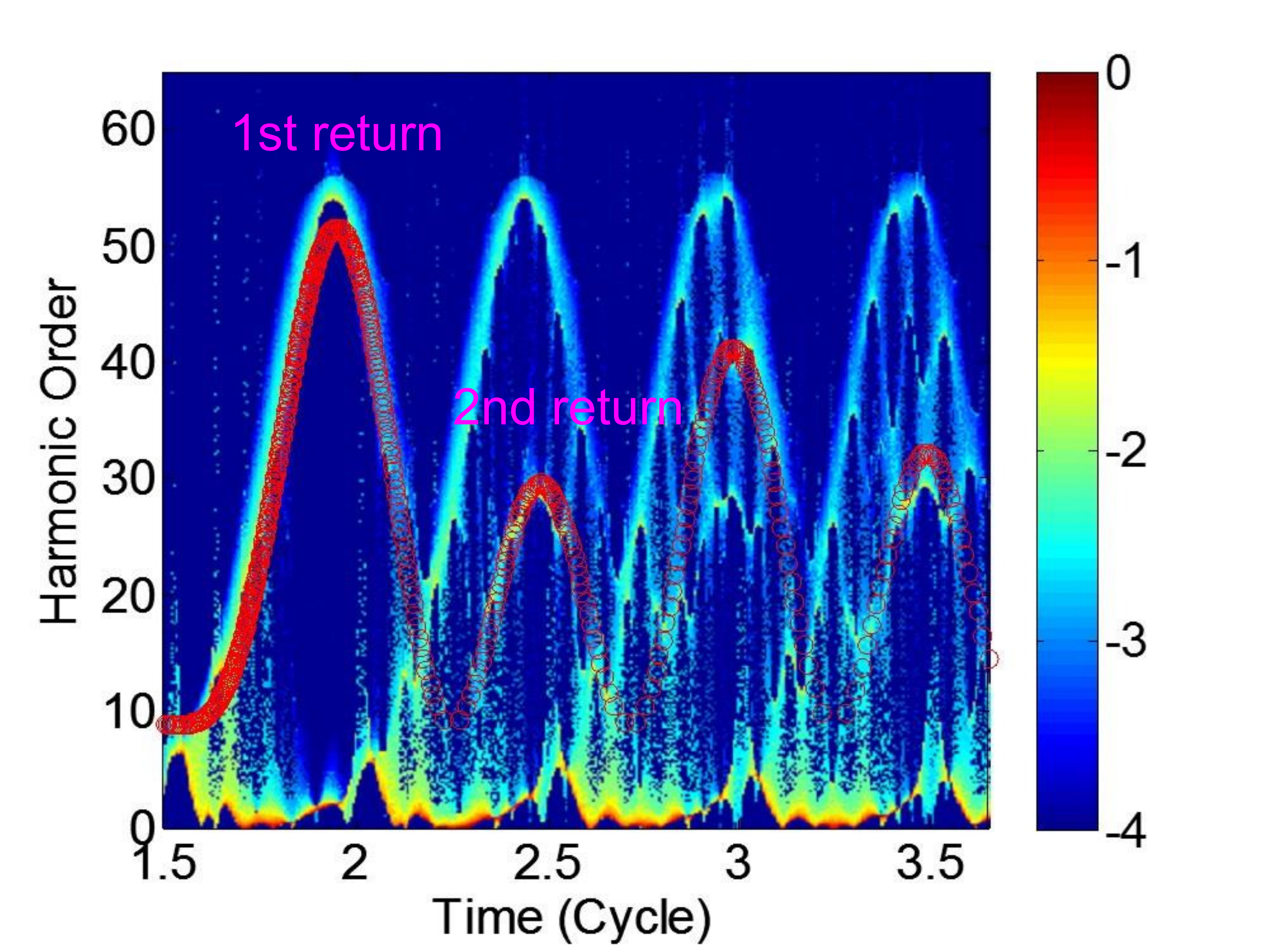}
  	\hspace*{3.0cm}(e)\hspace{6.5cm}(f)\\
 \caption{ TF representations in the tunneling ionization regime by the (a) RM-GT and (b) RM-SPWVD. The broadening artifacts caused by the window in both methods are eliminated, but it is difficult to remove the interference pattern in the SPWVD. TF representations in the tunneling ionization regime by the (c) SST-GT and (d) SST-MWT. The broadening artifacts caused by the window in both methods are alleviated.
(e) TF representation in the tunneling ionization regime by the second order SST-GT. Note that the diffusive pattern in the SST-GT is modified for the arches. (f) Comparison of the TF representation of the SST-GT and a trajectory of the electron released between the $1T$ and $1.5T$ (denoted by red circles) computed by the standard semiclassical approach suggested independently by Corkum \cite{Corkum} and Kulander {\itshape et al.} \cite{Kulander}.
 }
  \label{Fig7}
\end{figure*}

We further apply the RM techniques by reassigning the TF representations by the rule of centroids. In Fig.~\ref{Fig7}(a) and Fig.~\ref{Fig7}(b), we show the results of the RM-GT and the RM-SPWVD, representatives for the linear and quadratic type methods, respectively, can address the broadening issue in the original methods and provide distinct inner arches as well as other substructures.
Note that in Fig.~\ref{Fig7}(b), the arch in the duration of $2-2.5$T and around the 45th harmonic is an example of interference that is difficult to remove.
It is because that the RM techniques re-shuffle the original TF representation based on its local center of mass, and hence preserves the intrinsic artifact in the adopted transform.

By a reallocation rule that contains phase information of a linear type TF transform, the SST can enhance the TF representation by linear type TF transforms in a manner similar to that by the RM technique, as shown in Fig.~\ref{Fig7}(c) and (d).
Both the SST-GT and the SST-MWT can depict similar structure with appropriate the STFT and CWT parameters.
Note that as mentioned previously, the SST-GT inherited the fixed resolution feature in the STFT, and the SST-MWT maintains the adaptive resolution in the CWT.
By comparing Fig.~\ref{Fig7}(c) with Fig.~\ref{Fig7}(c), we see that the resolution in the RM-GT is more sharpen than that of the SST-GT.
One of the reasons is that while both temporal and frequency reassignments are considered in the conventional reassignment method, temporal reassignment is not taken into account in the SST (both SST-GT and SST-MWT).
In addition, the reallocation rule \eqn{sst_STFT.04} is only the first order approximation for the IF, resulting in diffusive pattern particularly for fast-varying chirp signal.
For the simulation in the multiphoton ionization regime, such diffusive pattern is negligible as the IF changes slowly. 
By using \eqn{sst_STFT.06}, a second-order reallocation rule for the SST-GT, the diffusive pattern can be further concentrated, as is shown in Fig.~\ref{Fig7}(e).

To shed some light on this TF representation for the tunneling ionization mechanism, 
we follow the standard semiclassical approach suggested independently by Corkum \cite{Corkum} and Kulander {\itshape et al.} \cite{Kulander}.
Here the electric-field force corresponding to the applied laser field in a.u. is  ${\bf F}_z=-E_0 E(t)\hat{z}$ , where $\hat{z}$ is the unit vector in the $z-$direction.
We assume the initial position and initial velocity is both zero \cite{Atto2}.

The trajectory of electrons released between $1T$ and $1.5T$ by using the extended semiclassical approach is denoted by red circles in Fig.~\ref{Fig7}(f).
This result is superimposed for the sake of comparison with the TF representation of the SST-GT.
The trajectory suggests that the largest arch in Fig.~\ref{Fig7} is associated with the first return time of the electron wave packet released in between $1T$ and $1.5T$ of the laser field,
and the smaller arch between $2T$ and $2.5T$, is the second return.


\begin{figure}
		\includegraphics[width=0.495\hsize]{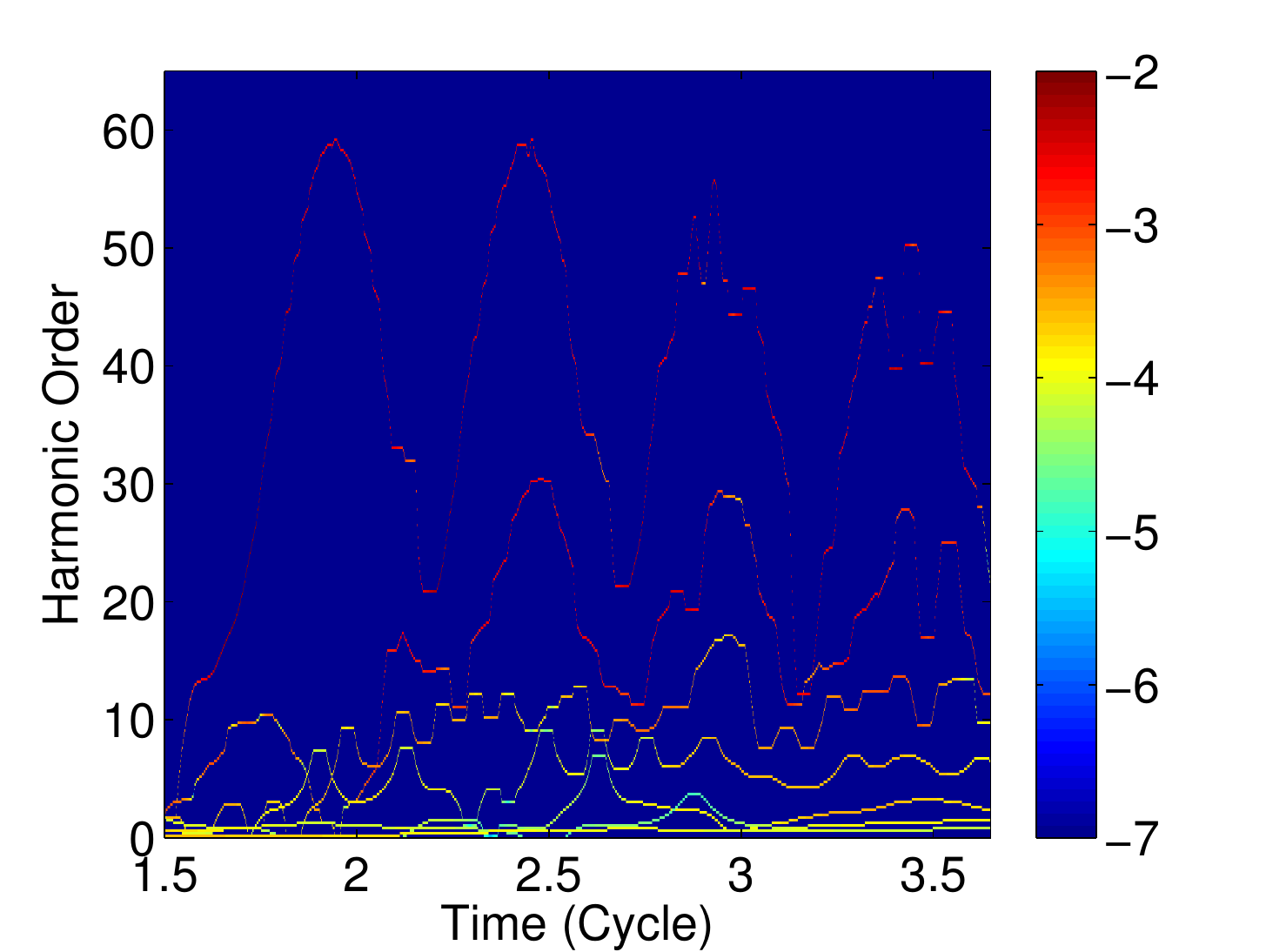}
		\includegraphics[width=0.495\hsize]{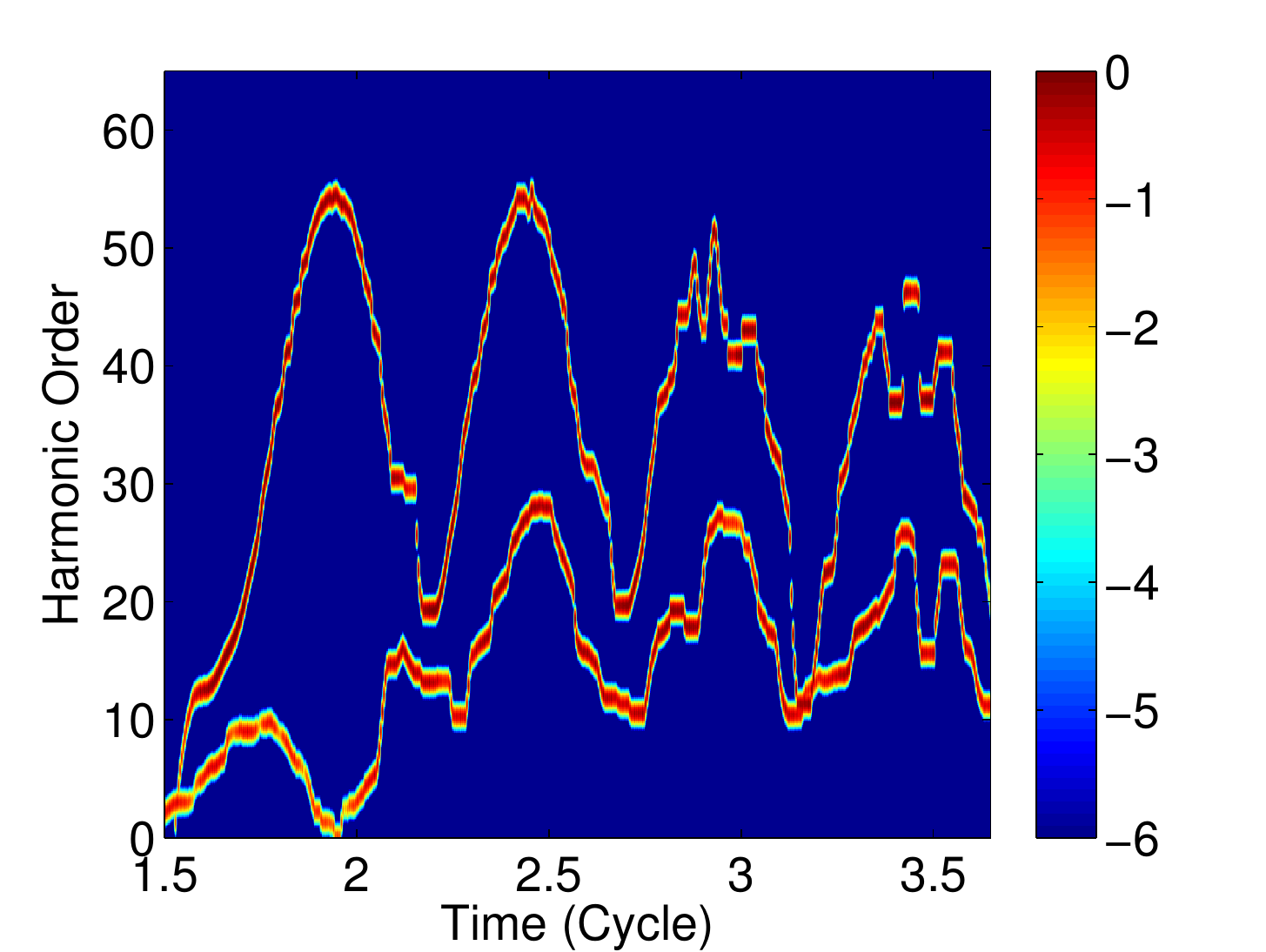}
  	\hspace*{1.3cm}(a)\hspace{6.0cm}(b)\\
\centering
  	\includegraphics[width=0.495\hsize]{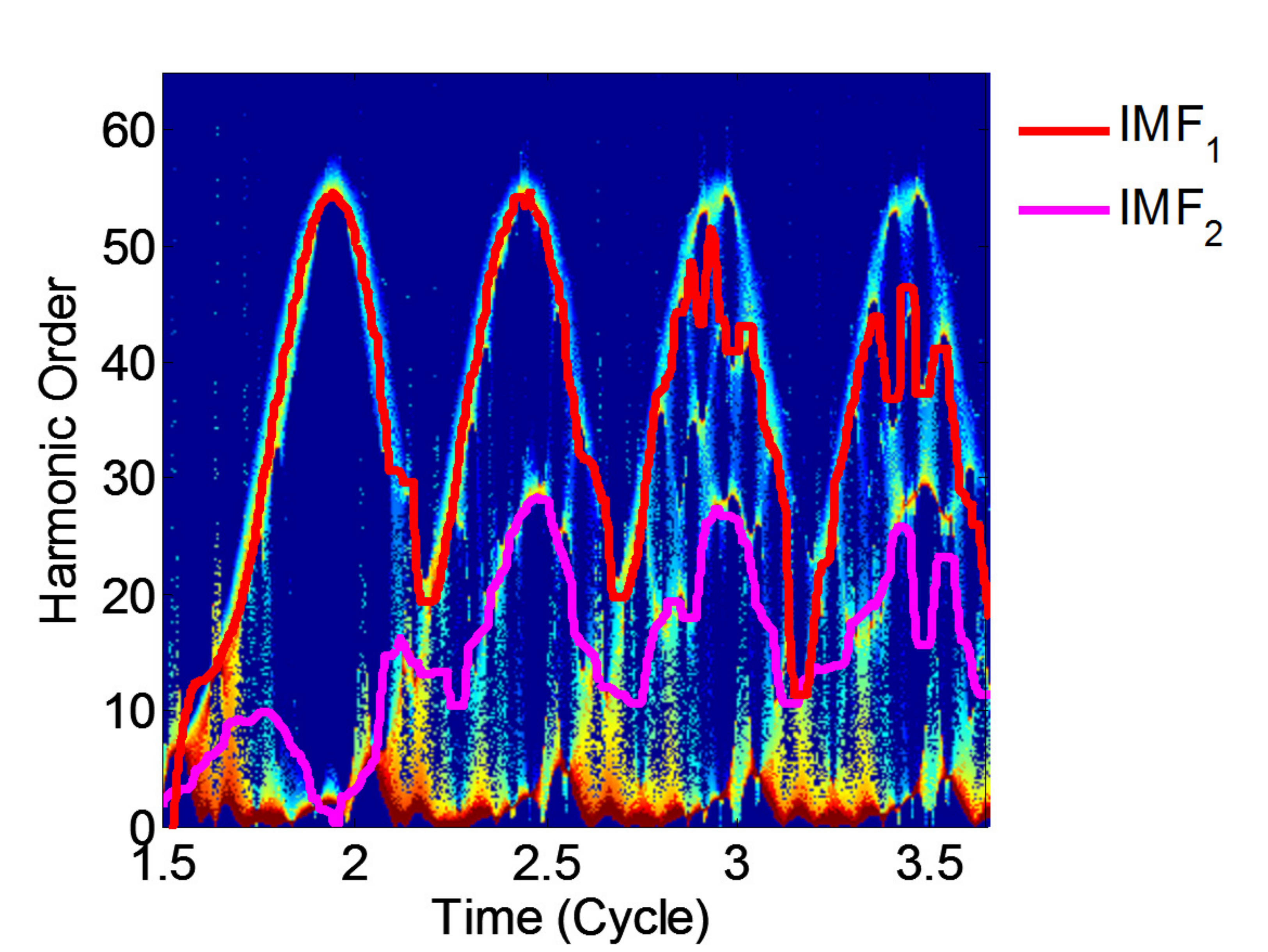}\\
  	\hspace*{0.0cm}(c)\\
  	\includegraphics[width=0.495\hsize]{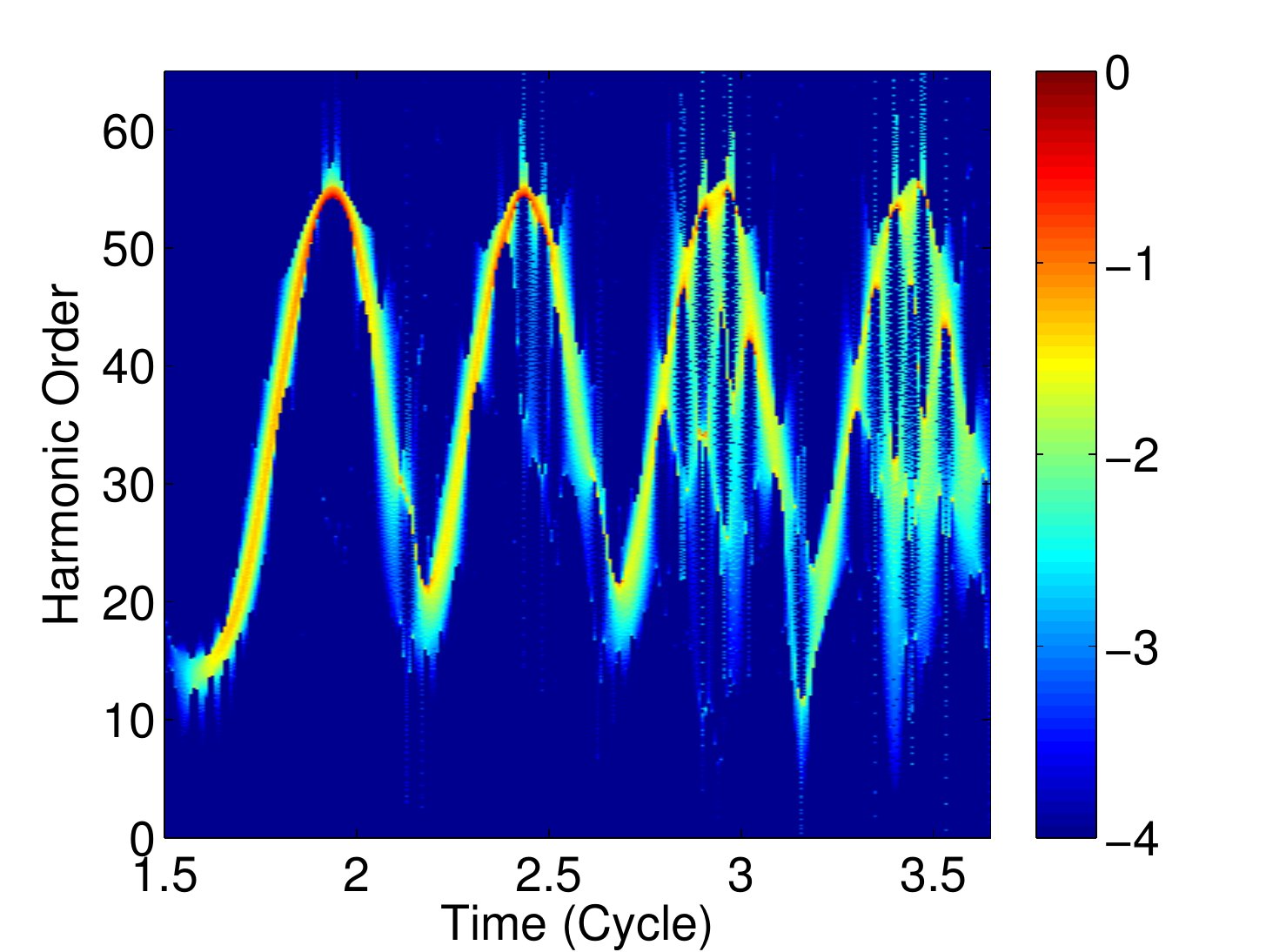}
  	\includegraphics[width=0.495\hsize]{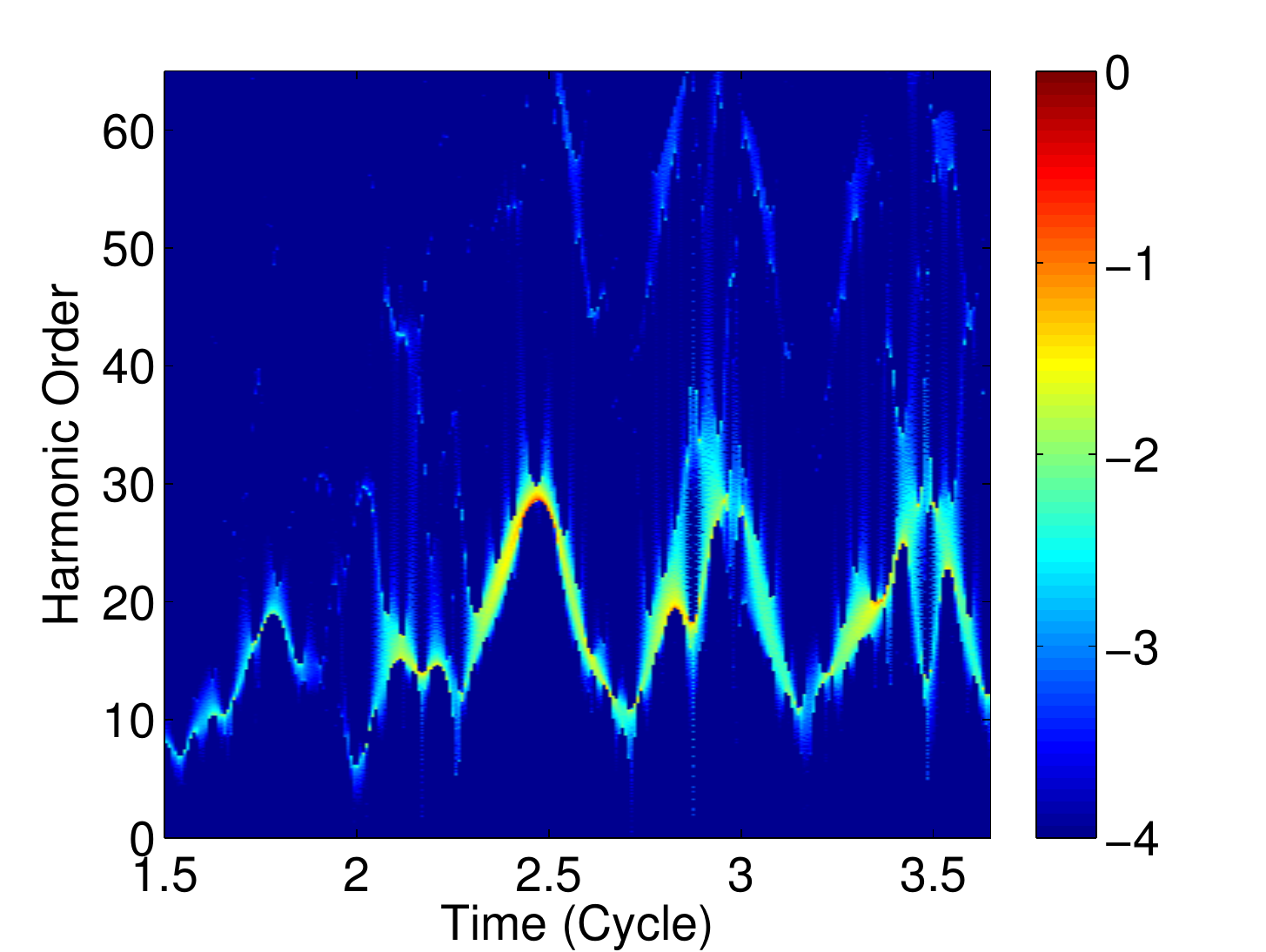}\\
  	\hspace*{0.cm}(d)\hspace{5.7cm}(e)\\
 \caption{ (a) TF representation in the tunneling ionization regime of the EMD-HS algorithm, which contains all IMF modes extracted by EMD. (b) TF representation in the tunneling ionization regime of the first two IMFs extracted by EMD. To enhance the visualization, the curves are thickened. (c) Comparison of the first two IMFs with the TF representation by the SST-GT. According to the semiclassical trajectory, the first IMF is the first return and the second IMF is the second return. (d) The first IMF and (e) the second IMF analyzed by the SST-GT. 
 }
  \label{emd}
\end{figure}


Although the SST depicts the first and multiple returns clearly, the sequence of these returns cannot be known.
To retrieve the first and second return quantitatively, Risoud {\itshape{et al.}} propose the application of the Hilbert transform on the dipole moment with the harmonics lower than the ionization potential filtered \cite{Risoud}. While there are more than one component simultaneously exist at each time, we consider the EMD scheme to decompose the filtered dipole moment.
(In this case, harmonics below the $10$th harmonic are filtered as they correspond to the bound part of the atomic spectrum.)
The stopping criterion in the EMD scheme is that the residue becomes monotonic. 
The HS for all modes of IMF are displayed in Fig.~\ref{emd}(a). Note that a median filter with a window length of $51$ points
is applied on all modes as the implementation of \eqn{emd.02} can introduce numerical errors. 
Despite there is no theoretical base for the IMFs by the EMD and the result is not consistent in the multiphoton ionization case, in this case we see that the first two modes can be associated with the first and second returns. 
Fig.~\ref{emd}(b) presents the HS for the first two IMFs and Fig.~\ref{emd}(c) compares the IMFs with the representation of the SST-GT. 
In Fig.~\ref{emd}(b), the IMFs are visually emphasized by additional Gaussian functions in both temporal and frequency domain.
Note that the IF revealed by the Hilbert transform is meaningful only if the IMF is consist of a single component, which is not guaranteed in the EMD. For example, for time greater than $2.5$ cycles, the HS of the first IMF is distorted because of more wave packet returning from previous emissions.

While the Hilbert transform is not theoretically suitable for this kind of oscillatory signal with time-varying AM and IF, we illustrate the IF of these decomposed modes by the SST-GT.
We employ the SST-GT on the first two IMFs, as presented in Fig.~\ref{emd}(d) and (e). Results in Fig.~\ref{emd}(d) and (e) suggest that expressing the IF of the IMF by the SST is more stable than using the Hilbert transform. 
Note that no median filter is necessary for each IMFs in the SST-GT analysis, which reduces the possible artifacts.

A summary for comparison of TF methods in the tunneling ionization regime is provided in Table $3$.


\begin{table}
\begin{threeparttable}
\label{list3}
\centering
\textbf{Table~3} Comparison of TF methods in the tunneling ionization regime \\[1ex]
\begin{tabular}{lllll}
Phenomenon& First Return  & Second Return &  Sequencing of    & Near and Below \\  
          &               &               &  Multiple Returns & Threshold Harmonics \\
\hline
GT      & Yes (blurred)       & Yes (blurred)   & No    &  No             \\
MWT       & Yes (blurred)       & Yes (blurred)   & No    &  Yes (blurred)  \\
WVD       & No                  & No              & No    &  No             \\
SPWVD     & Yes (blurred)       & No              & No    &  No             \\
RM-GT   & Yes                 & Yes             & No    &  Yes             \\
RM-SPWVD  & Yes                 & No              & No    &  No             \\
SST-GT  & Yes (diffused)      & Yes             & No    &  No      	\\
$2$nd order SST-STFT	& Yes     & Yes             & No    &  No        \\	
SST-MWT   & Yes (diffused)      & Yes             & No    & Yes       	\\		
EMD-HS    & Yes (mode mixing)   & Yes (mode mixing) & Yes & No          \\
EMD-SST   & Yes (mode mixing)   & Yes (mode mixing) & Yes & No          \\
\end{tabular}
\end{threeparttable}
\end{table}


\section{Conclusions}\label{Section:Conclusion}

This paper provides a benchmark of analyzing time-dependent quantum systems by applying several contemporary TF methods.
Within the same physical model and computational scheme, different features of TF representations provided by different TF methods may lead to conflicting interpretation in physics.
In the multiphoton ionization regime, linear TF methods with an appropriate window function, can vaguely depict discrete odd harmonics in the HHG process, yet the detail features such as the AC Stark effect cannot be revealed.
In addition, in the tunneling ionization regime, the first-return trajectories are roughly illustrated, but the multiple returns are indistinct because of the broadening induced by the window.
While present little broadening, TF representations by the WVD suffer from artificial interference patterns and lead to incorrect interpretation of physical mechanisms, such as even harmonics in the multiphoton ionization regime and false trajectories in the tunneling ionization regime.
To eliminate the artifacts mentioned above, several modern TF analysis methods are introduced in quantum dynamics for the first time.
However, Cohen class distributions such as the SPWVD can only remove portion of interference patterns.
Reassignment techniques and the SST can depict accurate dynamic process in the two regimes based on the principle of local centroid and instantaneous frequency, respectively.
However, there is no inverse transform in the reassignment techniques.
On the other hand, the SST preserves signal causality and allows mode reconstruction.
We demonstrate that after removing the broadening, the AC Stark effect in the multiphoton ionization regime and the multiple returns in the tunneling ionization regime are revealed.
Finally, we relate the IMFs from the EMD to the multiscattering of the electron energy distribution in the tunneling ionization regime.

We believe that in addition to the atomic hydrogen system, the contemporary TF methods have potential for exploring other complicated quantum systems. 
In case of analyzing dynamics in an unknown quantum system, this paper suggests the following procedure.
First, features obtained by the WVD can be used to determine the parameter for the linear type methods.
Then the SST can be applied to obtain TF representations for further interpretation.
We hope that this research can serve as a cornerstone in applications of TF analysis for fields including but not limited to, attosecond physics, nuclear magnetic resonance, ultrafast dynamics in atoms and molecules.

\section{Acknowledgments}

The authors would like to thank Dr. Elise Y. Li and Yu-lin Sheu for their help in servers.

\appendix

\section{Adaptive harmonic model}

Fix constants $0\leq \epsilon\ll 1$, $0<d<1$, $c_2>c_1>\epsilon$ and $c_2>c_3>\epsilon$. Consider the functional set $\mathcal{Q}_{\epsilon}^{c_1,c_2,c_3}$, which consists of functions $x(t) = A (t)\cos(2\pi\phi(t))\in C^1(\mathbb{R})\cap L^\infty(\mathbb{R})$
where the following conditions hold:
\begin{equation*}
\left\{\begin{array}{l}\vspace{.2cm} 
A\in C^1(\mathbb{R})\cap L^\infty(\mathbb{R}),\quad\phi\in C^3(\mathbb{R}),\\ \vspace{.2cm} 
\inf_{t\in\mathbb{R}} A(t)\geq c_1,\quad \inf_{t\in\mathbb{R}}\phi'(t)\geq c_1,\\ \vspace{.2cm} 
\sup_{t\in\mathbb{R}} A(t)\leq c_2,\quad\sup_{t\in\mathbb{R}}\phi'(t)\leq c_2,\quad\sup_{t\in\mathbb{R}}|\phi''(t)|\leq c_3,\\ \vspace{.2cm} 
|A'(t)|\leq \epsilon\phi'(t),\quad |\phi'''(t)|\leq \epsilon\phi'(t) \quad\mbox{ for all }t\in\mathbb{R},
\end{array}\right.
\end{equation*}
We call $x= A (t)\cos(2\pi\phi(t))\in \mathcal{Q}_{\epsilon}^{c_1,c_2,c_3}$ an {\it intrinsic mode type} (IMT) function,
$\phi(t)$ the phase function, the derivative of the phase function $\phi'_k (t)$ is called the \textit{instantaneous frequency} (IF) and $A(t)$ is called the {\it amplitude modulation}. When $|\phi''(t)|\ll \epsilon \phi'(t)$ for all $t\in\mathbb{R}$, we call the signal $x$ oscillatory with {\it slowly varying IF}; otherwise we call it oscillatory with {\it fast varying IF}. Locally an IMT function with slowly varying IF behaves like a harmonic function and an IMT function with fast varying IF behaves like a linear chip function. IMT function serves as a mathematical formula for the IMF considered in the EMD algorithm, i.e., the IMT functions contain the properties defined in the IMF, but the reverse inclusion does not hold.

To describe an oscillatory signal, the following adaptive harmonic model is commonly considered in the time-frequency analysis literature \cite{SST1,SST2,SST3}.
Fix constants $0\leq \epsilon\ll 1$, $d>0$ and $c_2>c_1>0$. Consider the functional set $\mathcal{Q}_{\epsilon,d}^{c_1,c_2,c_3}$, which consists of functions in $C^1(\mathbb{R})\cap L^\infty(\mathbb{R})$ with the following format: 
$x(t) = \sum_{\ell=1}^K x_k(t)$, where $K$ is finite and $x_k(t)=A_\ell(t)\cos(2\pi\phi_\ell(t))\in \mathcal{Q}_{\epsilon}^{c_1,c_2,c_3}$. Let $\phi'_k(t)>\phi'_{k-1}(t)$. In the STFT-SST, successive IMTs for all time $t$ have to be separated by at least $d$, i.e., $\phi'_k(t)-\phi'_{k-1} (t)>d$.

\section{The Floquet Method}

The total Hamiltonian of atomic hydrogen can be expressed as $H(t)=H_0+H_1(t)$, where $H_0$ is the unperturbed hydrogen Hamiltonian and $H_1(t)=-zE(t)$ describes the laser-atom interaction. 
In the time-dependent generalized pseudospectral simulation, the time period $T$ is fixed but the field amplitude varies. To calculate the shifting of orbital energies, we separate each cycle and take its field amplitude, and then assume the hydrogen in a laser field with the particular constant amplitude. For example, for the third cycle in Fig.~\ref{Fig1}(a), the maximal amplitude is $4.86\times 10^{-4}$, and we assume hydrogen in the laser field with the constant amplitude.

When the Hamiltonian is periodic with a time period $T$, i.e., $H(t)=H(t+T)$, we can apply the Floquet theory \cite{Shirley} and derive a time-independent infinite-dimensional eigenvalue matrix equation \cite{Hsu} as follows:
\begin{eqnarray}
\label{HF1}
\sum\limits_{nlmk}\langle \langle nlmk| \widehat{H}_F |n'l'm'k'\rangle\rangle \phi^{n'l'm'k'}_{\lambda\xi}
=\varepsilon_{\lambda\xi}\phi^{nlmk}_{\lambda\xi}, \label{app.b1}
\end{eqnarray}
Here $\langle \langle nlmk| \widehat{H}_F |n'l'm'k'\rangle\rangle$ denotes the matrix elements of the time-averaged Floquet Hamiltonian over a period $T$, where the outer bra-ket notation refers to the inner product over $t$.
We use the orbitals of an unperturbed hydrogen as a basis set since we only consider a weak field regime.
As a result, $n$, $l$, and $m$ represent the principal quantum number, angular momentum quantum number, and magnetic quantum number of hydrogen, respectively. $k$ is an integer.

The matrix element in Eq.~(\ref{HF1}) can be evaluated as
\begin{eqnarray}
&&\langle\langle nlmk| \widehat{H}_F |n'l'm'k'\rangle\rangle  \nonumber \\
&=& E_{nlm}\delta_{nn'}\delta_{ll'}\delta_{mm'}\delta_{kk'}  \nonumber \\
&-& \langle nlm| z |n'l'm'\rangle \frac{eE_0}{2}(\delta_{k,k'+1}+\delta_{k,k'-1}) \nonumber \\
&+&k\hbar\omega\delta_{nn'}\delta_{ll'}\delta_{mm'}\delta_{kk'}
\label{app.b2}
\end{eqnarray}

The values of $\langle nlm| z |n'l'm'\rangle$ can be computed analytically.
For examples, in atomic units, $\langle 210| z |1'0'0'\rangle=0.7449$, $\langle 310| z |1'0'0'\rangle=0.2983$, $\langle 210| z |3'0'0'\rangle=0.5418$, and $\langle 310| z |2'0'0'\rangle=1.7695$.

\end{document}